\documentclass{aastex631}

\usepackage{multirow}
\begin{document}

\title{Exocomet models in transit: light curve morphology in the optical -- near infrared
wavelength range}

\author[0000-0003-3754-7889]{Szil\'ard K\'alm\'an}
\affiliation{Konkoly Observatory, HUN-REN Research Centre for Astronomy and Earth Sciences, Konkoly Thege 15-17, 1121 Budapest, Hungary}
\affiliation{HUN-REN CSFK, MTA Centre of Excellence, Budapest, Hungary}
\affiliation{HUN-REN-ELTE Exoplanet Research Group, Szombathely, Szent Imre h. u. 112., H-9700, Hungary}
\affiliation{ELTE E{\"o}tv{\"o}s Lor\'and University, Doctoral School of Physics,  Budapest, Pázmány Péter sétány 1/A, H-1117, Hungary}
\author[0000-0002-0606-7930]{Gyula M. Szab\'o}
\affiliation{HUN-REN-ELTE Exoplanet Research Group, Szombathely, Szent Imre h. u. 112., H-9700, Hungary}
\affiliation{ELTE E{\"o}tv{\"o}s Lor\'and University, Gothard Astrophysical Observatory, Szombathely, Szent Imre h. u. 112., H-9700, Hungary}

\author[0000-0002-8722-6875]{Csaba Kiss}
\affiliation{Konkoly Observatory, HUN-REN Research Centre for Astronomy and Earth Sciences, Konkoly Thege 15-17, 1121 Budapest, Hungary}
\affiliation{HUN-REN CSFK, MTA Centre of Excellence, Budapest, Hungary}
\affiliation{Institute of Physics and Astronomy, ELTE Eötvös Loránd University,
Budapest, Hungary}
\correspondingauthor{Szil\'ard K\'alm\'an}
\email{xilard1@gothard.hu}



\begin{abstract}
Following the widespread practice of exoplanetary transit simulations, various presumed components of an extrasolar system can be examined in numerically simulated transits, including exomoons, rings around planets, and the deformation of exoplanets. Template signals can then be used to efficiently search for light curve features that mark specific phenomena in the data, and they also provide a basis for feasibility studies of instruments and search programs. In this paper, we present a method for exocomet transit light curve calculations using arbitrary dust distributions in transit. The calculations, spanning four distinct materials (carbon, graphite, pyroxene, and olivine), dust grain sizes ($100$\,nm -- $300$\,nm, $300$\,nm -- $1000$\,nm, and $1000$\,nm -- $3000$\,nm) encompass light curves in VRJHKL bands. We also investigated the behavior of scattering colors. We show that multicolor photometric observations are highly effective tools in the detection and characterization of exocomet transits. They provide information on the dust distribution of the comet (encoded in the light curve shape), while the color information itself can reveal the particle size change and material composition of the transiting material, in relation to the surrounding environment. We also show that the typical cometary tail can result in the wavelength dependence of the transit timing. We demonstrate that multi-wavelength observations can yield compelling evidence for the presence of exocomets in real observations.
\end{abstract}


\section{Introduction} \label{sec:intro}
Comets, mostly known as spectacular sojourners on our skies \citep{2017come.book.....K}, have been counted for a long time among the most ancient reservoirs of solid material and ice in the Solar System. As witnesses of the formation of the outer Solar System \citep[e.g.][]{1963Icar....2..396D,1984SSRv...38...35E,2013AstSR...9..146V}, their orbits can evolve through gravitational perturbations, causing them to migrate to the inner Solar System -- in a process that presumably happens very similarly in the extrasolar systems as well \citep[e.g.][]{2007ApJ...654..595G,2017MNRAS.464.1415M}. Comets are believed to be one of the possible sources of water and nitrogen \citep[e.g.][]{2012E&PSL.313...56M,2015Sci...347A.387A,2020Sci...367.7462P} and possibly even organic material on Earth \citep[e.g.][]{2016Natur.538...72F,2021PhLRv..37...65D}. They are therefore invoked as essentials for the volatiles and habitability of Earth and, based on this analogy, also on other planetary systems \citep[e.g.][]{2018MNRAS.479.2649K}. Thus, these objects are also important from an astrobiological perspective \citep{2019MmSAI..90..581W}. 

Compared to the $\sim 4150$ known exoplanetary systems at the time of writing, there are only a few systems in which exocometary activity has been detected. There are two main approaches to the detection of extrasolar comets: photometry and spectroscopy \citep{2021KPCB...37...64P}. In the circumstellar disk of the $\beta$ Pictoris system, \cite{1987A&A...185..267F} detected time-variable features in Ca \textsc{ii} H \& K spectral lines, which are attributed to the so-called Falling Evaporating Bodies \citep[FEBs][]{1994Ap&SS.212..147B, 1989A&A...215L...5L, 2000Icar..143..170B}. Given that FEBs were first detected and are perhaps the most studied in the $\beta$ Pictoris system \citep{2007A&A...466..201B, 2019MNRAS.489..574T}, these spectroscopic transients are also known as $\beta$ Pic-like phenomena. \cite{2014Natur.514..462K} identified 493 individual exocomets in $\beta$ Pictoris through the study of these features. The large number of comets detected allowed for a clustering of these objects, yielding two groups of distinct origins and properties \citep{2014Natur.514..462K, 2014Natur.514..440R}. Observations of FEBs are not limited to the $\beta$ Pic system. 
\cite{2007ApJ...656L..97R} detected time-dependent absorption features in the Na \textsc{i} D doublet of HD 32297, which is related to the debris disk surrounding the star. Other well-known examples include HD 172555 \citep{2014A&A...561L..10K, 2018AJ....155..242G}, HD 100546 \citep{1997ApJ...483..449G}, HD 24966, HD 38056, HD 79469 and HD 225200 \citep{2018MNRAS.474.1515W}. As a further step in interpreting these spectroscopic observations, \cite{2018MNRAS.479.1997K} proposed a method to retrieve some of the orbital elements of the exocomets in question. \cite{2020A&A...639A..11R} carried out a large spectroscopic survey of (suspected) exocomet host stars, finding that $17\%$ of stars showing near infra-red excess allow for a detection of FEB-like events, suggesting a weak link between the evolution of a protoplanetary disk and the features attributed to exocomets.

The FEB-detections listed above rely on spectroscopic time series observations, which are complicated by nature. \cite{1999A&A...343..916L} proposed the charateristic ``round triangle'' shape of exocometary transits, which are thought to be photometric counterparts of the FEBs \citep{2019A&A...625L..13Z}, and subsequently constructed a library of various transit shapes \citep{1999A&AS..140...15L}. \cite{1999A&A...343..916L} present a light curve model constructed of two components: extinction and forward scattering, both of which are based on physical modeling of cometary activity. The total contribution of these effects depends strongly on the grain size of the dust (and gas) emitted from the comet as well as on its dust production rate. The shapes of the transit light curves that result are colloquially called ``sharkfins''. These transit events are naturally detectable in long baseline datasets of survey-type space observatories, such as \texttt{Kepler} \citep{2010Sci...327..977B}, \texttt{TESS} \citep[Transiting Exoplanet Survey Satellite]{2015JATIS...1a4003R} or the upcoming \texttt{PLATO} \citep[PLAnetary Transits and Oscillations of stars]{2014ExA....38..249R}. \texttt{TESS} also revealed several hundred so-called dippers \citep{Tajiri2020}, which are generally young stars characterized by transients with variable depth and lasting characteristically for 0.5--1 days. These transients are thought to have multiple origins, from the disk warps and accretion columns of T tauri stars (represented by rather periodic signals) to the transit of giant exocomets in the line of sight (represented by aperiodic components). \cite{2020PASP..132j1001S} also discussed the possibility of interstellar comets as the origin of the signals which would naturally be single-occurance events. Although the low rate of periodicity in a system is a diagnostic of the presence of exocomets, there is still a need for independent evidence for the cometary origin \citep{Tajiri2020,Gaidos2022a,Gaidos2022b}.


The detection of exocomet transits \citep{2018MNRAS.474.1453R,2019A&A...625L..13Z, 2019MNRAS.482.5587K,2022NatSR..12.5855L, 2023A&A...671A..25K} often relies on pattern search methods\footnote{Private communication with S. Zieba} \citep[including the box-fitting approach of ][that has to include a false-positive rejection stop due to exoplanets, binary stars etc.]{2019MNRAS.482.5587K}, which are known to be sensitive only to patterns that were used for training. In case of the box-fitting approach, this sensitivity arises at the false-positive rejection step. Most ``exocomet transit'' potometric transients indeed represent a sharkfin-like feature, but these are the features the methods are looking for. With a sophistically trained pattern generator, we can derive template light curves of transiting exocomets that we can include in the actual search and interpretation efforts.

In this paper, we utilize a simple model to generate synthetic exocometary transit light curves. We put comet-like model distributions into transit and calculate the radiation transfer as a function of the angle of the incident backlight. We present template light curves from these transit simulations from optical to near-infrared (NIR) wavelengths. We also explore the ``color of comet dust'' \citep[e.g.][]{2001JGR...10610113K,2007P&SS...55.1010L} when transiting a star, comparing the effect of dust size and material dependence along the examined wavelengths. 

The paper is structured as follows. In Sect. \ref{sec:methods} we describe the exocomet models, as well as the light curve simulations (including both extinction and forward scattering). In Sect. \ref{sec:res}, we present a library of light curves that are made publicly available. In Sect. \ref{sec:disc}, we place our models into context and consider the possible implications regarding the future of the hunt for exocomets.

\section{Methods} \label{sec:methods}

\subsection{Radiative transfer modeling}

\begin{figure}
    \centering
    \includegraphics[width = 0.3\textwidth]{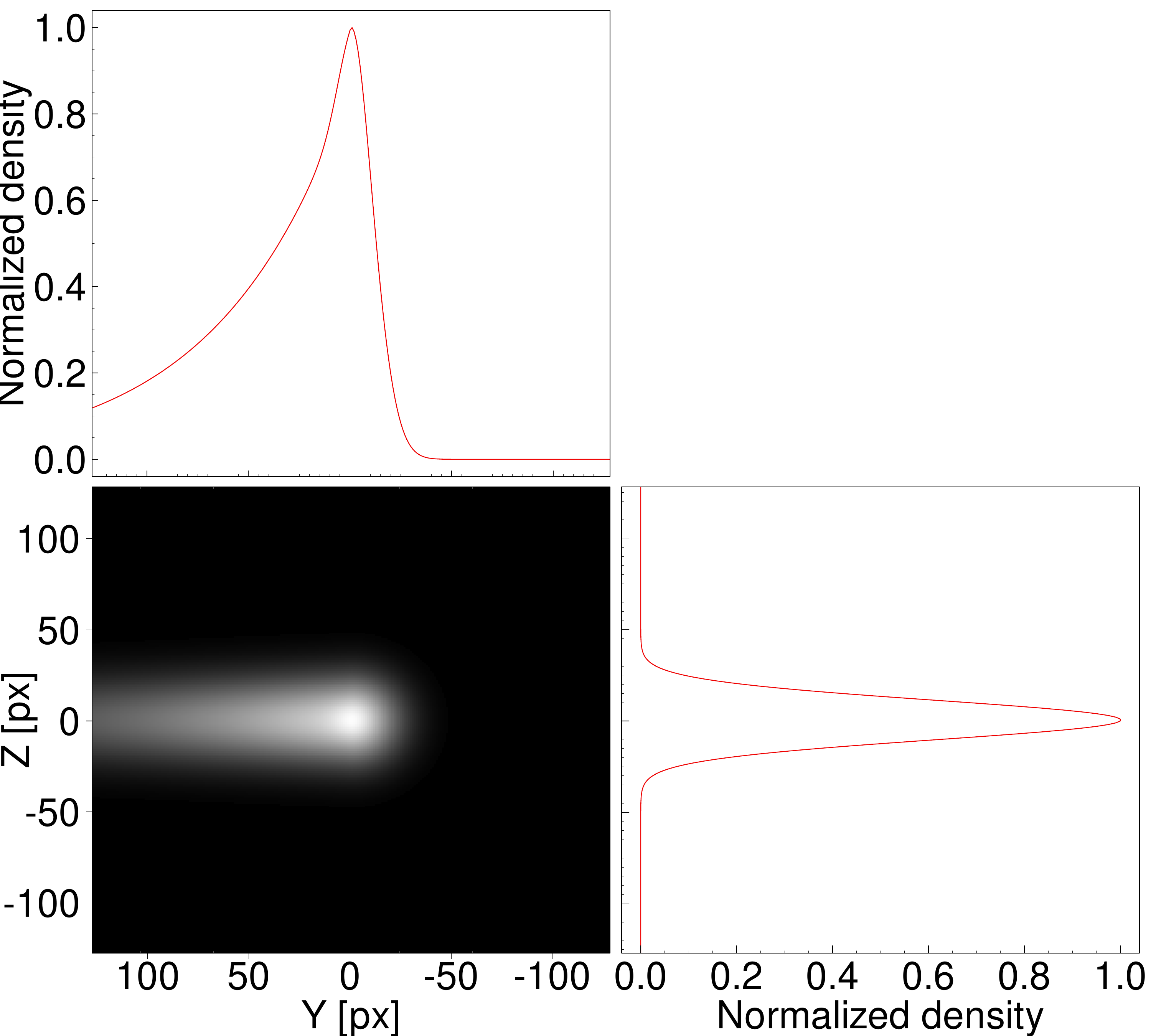}
    \includegraphics[width = 0.3\textwidth]{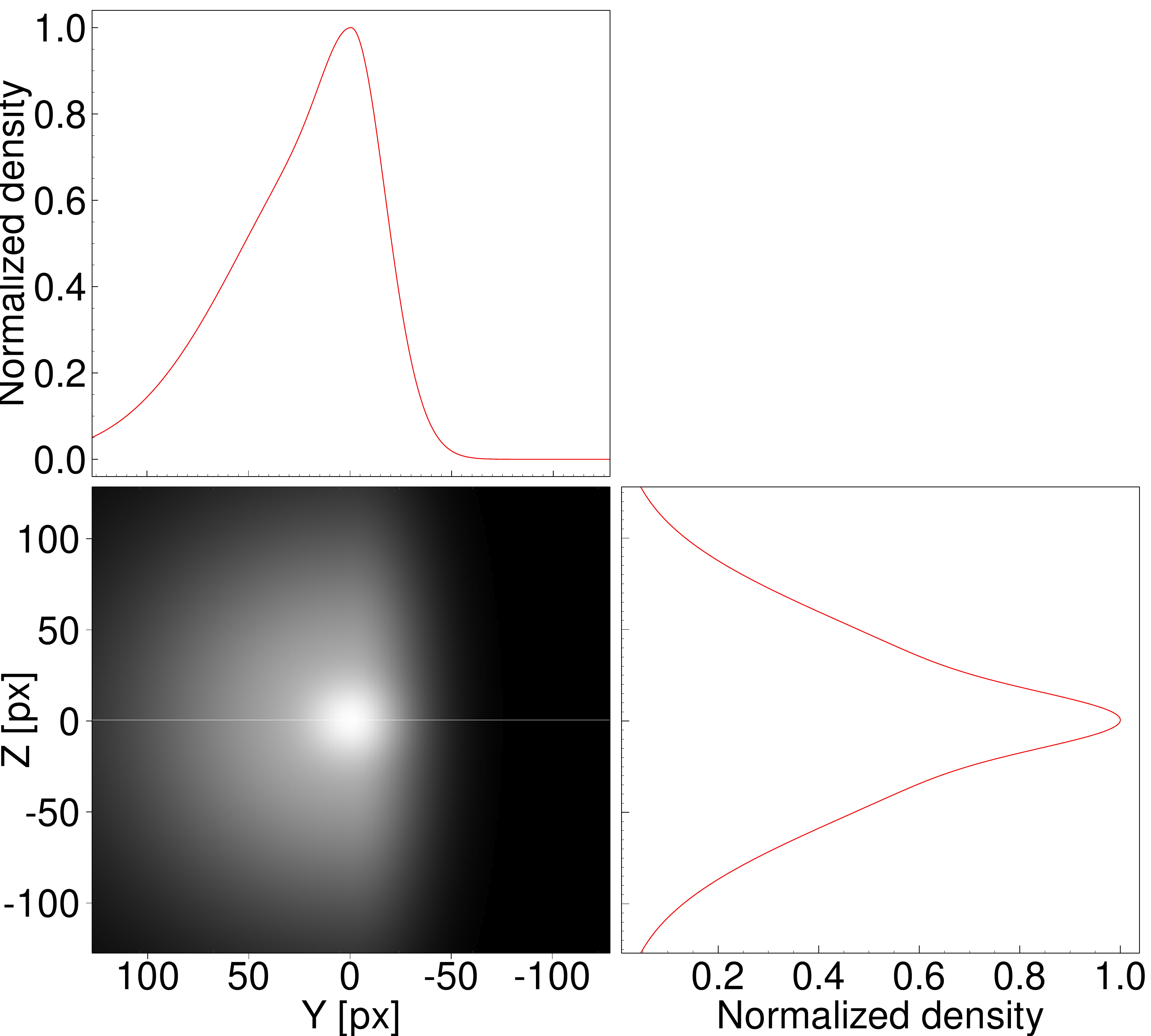}
 \includegraphics[width = 0.3\textwidth]{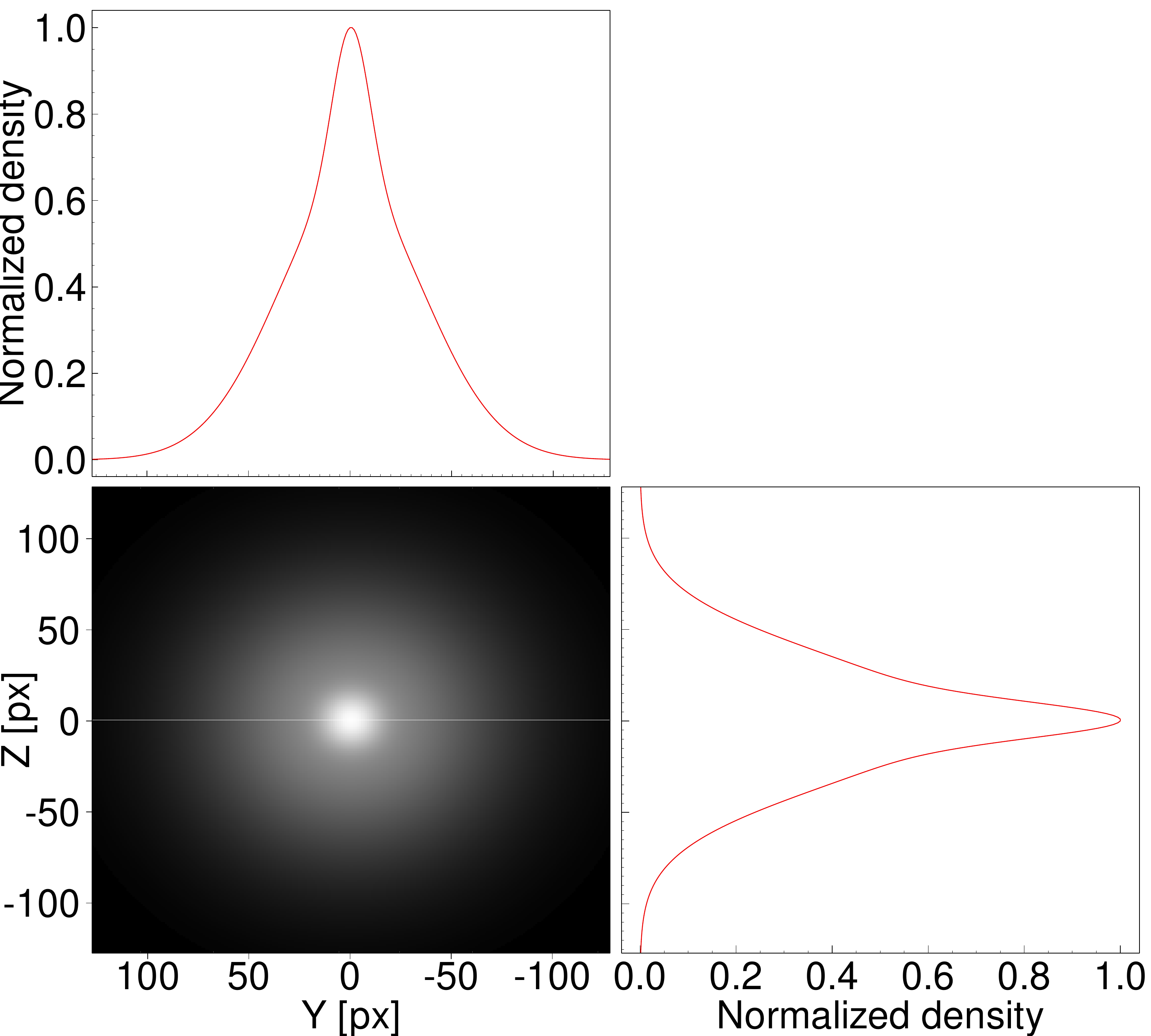}
    \caption{Density distribution of the comet models in this study. The figure panels show the comet with a narrow tail, the comet with a fan-like tail and the comet without a tail. The 2D images show the comet models as seen in reflected light in sky projection, together with the projected densities on the y and z axis (top and bottom right panels).}
    \label{fig:distributions}
\end{figure}

Radiative transfer modeling of dust structures, assumed to be responsible for the obscuration of stellar light in transient phenomena, is performed using RADMC-3D \citep{radmc}. RADMC-3D is a freely available open source code package for diagnostic radiative transfer calculations. It calculates, for a given geometrical distribution of gas and/or dust, what its images and/or spectra look like when viewed from a certain angle, including optical depth calculations, both in scattered light and in thermal emission. To perform exocomet light curve synthesis, we use RADMC-3D 2.0 and a volume of 256$^3$ cells in Cartesian coordinates, with a box edge size adjusted to the actual density distribution (see Sect.~\ref{sect:dd}), typically in the order of 0.05\,au. The center of the coordinate system is the center of the simulation volume. The central star is located at a distance of $r_h$ from the center of the coordinate system on the x-axis, outside the simulation volume. We considered two cases: a `solar-like' star (with a photospheric temperature of 5780\,K a radius of $R_\odot = 6.96\times10^5$\,km and $\log g = 4,5$) and, since most exocomet detections have been made around A stars \citep[e.g. the $\beta$ Pic system, ][]{2019A&A...625L..13Z, 2022NatSR..12.5855L}, a star with a photospheric temperature of $8200$\,K, $R_\star = 1.7 R_\odot$ and $\log g = 4.0$. Although stellar luminosity is known to greatly influence the formation of cometary comae and tails \citep{1999A&AS..140...15L}, we used the same dust distributions for both stars.

The materials used for radiative transfer modeling are characterized by their composition and size distribution, as explained below. We utilize the {\sl optool}  package \citep{optool} to obtain the complex dust particle opacities, using materials from {\sl optool}'s material library, and assuming a specific grain size distribution. We consider different types of materials for our dust structures in the different simulations, but we assume that the whole simulation volume is homogeneous material-wise. Anisotropic scattering is treated by applying the Henyey-Greenstein function (option {\sl scattering\_mode\_max\,=\,2} in RADMC-3D), and using the scattering opacity and $g$ anisotropy parameters obtained for the specific material with {\sl optool}. To be able to disentangle the effect of grain size, we use three size distributions. The number of particles ($N$) within these ranges is given by
\begin{equation}
    N \propto a^{-q}.
\end{equation}
We set the size distribution exponent $q = 3.5$ in all cases \citep{1969JGR....74.2531D}. The size ranges of the three distributions are a\,=\,0.1--0.3\,$\mu$m (labeled as 200\,nm and corresponding to a mean size of $143.82$~nm), 0.3--1.0\,$\mu$m (650\,nm, with a mean size of $439.51$~nm) and 1.0--3.0\,$\mu$m (2\,$\mu$m), which corresponds to a mean particle size of $1438.2$~nm. We selected these specific grain size intervals to sample the total scattering cross section provided by \cite{2024MNRAS.527.3559P}, at the lower size ranges of that distribution, as well as the range where the rapid changes in the total cross section occur. Larger grains with a\,$\gtrsim$\,10\,$\mu$m cause proportionally smaller extinction and increased brightness from forward scattering and are not considered in our present analysis. 
While in a real comet we likely see a mixture of different materials, our goal here is to see the effect of a single composition separately. Therefore, in our calculations, we use four different homogeneous compositions, in each case using each of the following material compositions separately: graphite \citep{Draine2003}, amorphous pyroxene with 70\% Mg content \citep{Dorschner1995,Henning1996}, amorphous olivine with 50\% Mg content \citep{Dorschner1995,Henning1996}, and carbon grains \citep{Zubko1996}. While the choice of materials is arbitrary, we note that the carbonaceous \citep{2021PSJ.....2...25W} and silicate dust \citep{1999ApJ...517.1034W} is common in the comets of the Solar System. The debris disk of $\beta$ Pictoris is also known to host various silicates \citep{2004Natur.431..660O, 2022ApJ...933...54L, 2024AJ....167...69R}. In the case of a more realistic scenario when dust grains with different compositions are mixed, the total extinction / amount of scattered light can be obtained from the linear combination of the optical depths / scattered light contribution of the individual materials, assuming that the optical depth of the overall dust structure remains low ($\tau$\,$\ll$\,1) and the radiation field is dominated by the radiation of the central star even inside the dust structure. As we will see, this is the case in our simulations. 
In case of Trans-Neptunian objects (in a frozen environment) the model spectra usually assume the presence of compound materials such as ices depleted to solid particles \citep[e.g.][]{2018AJ....155..170S}. However, in case of comet Hale-Bopp, the spectra are well described as a sum of contributions of simple materials \citep[pyroxene and olivine in particular][]{1999ApJ...517.1034W}. The lack of ice layer on comet dust in general seems to be plausible taken the low ice-to-dust ratio on comet layers into account, and the energetic irradiation field. For this reason, we restrict our simulations to homogeneous materials.

Extinction and forward--scattering light curves are calculated separately, assuming that the dust distribution (i.e. the comet) passes in the line of sight of the central star, orbiting it at a specific distance from the star (0.1\,au on a circular orbit), the orbital plane is seen perfectly edge-on, and the occultation is central. We therefore neglect the impact of the orbital inclination (relative to the line of sight), which is known to influence the exact shape and depth of the transits \citep{1999A&AS..140...15L}.

For each light curve position, the central star is at a different location behind the density distribution (simulation volume). We cover the whole occultation event by changing the observer's aspect angle by discreet intervals ($0.1^\circ$ during the transit itself, $0.2^\circ$--$0.05^\circ$ out-of-transit). This is equivalent to the motion of the comet in front of the central star. The extinction light curve is obtained from the optical depth map, which is calculated by RADMC-3D for the specific observing geometry, convolved with the limb-darkened, resolved disk of the central star. We utilized the quadratic l imb-darkening rule \citep{1985A&AS...60..471W}, where the surface intensity of the star can be expressed as
\begin{equation}
    \frac{I \left( \gamma \right)}{I(0)} = 1 - u_1 \cdot \left(1-\cos\left(\gamma \right) \right) - u_2 \cdot \left( 1-\cos \left( \gamma \right) \right)^2.
\end{equation}
Here, $\gamma$ is the angle between the normal vector on the stellar surface and the line of sight. We adopted the coefficients $u_1$ and $u_2$ from \cite{1998A&A...335..647C}, these are listed in Table \ref{tab:ldc}. 

\subsection{Light curve calculations}\label{sect:lc_calc}
The forward-scattered light curve component corresponds to the total intensity of the surface brightness distribution of the same observing geometry, without the contribution of the central star (also directly obtained with RADMC-3D). It is calculated for a point source (assuming blackbody radiation), thus we neglect the effect of limb-darkening on this component. It can easily be scaled between the two stellar types (Solar-like and A-type star) by calculating the ratio of the emitted photons in each star. In this case, the A star we selected emits $4.1\times$ as many photons (at every wavelength), yielding $4.1\times$ higher amplitudes for this light curve component, as we show below. As a final step, we fit a spline function to interpolate both components of the light curves.

\begin{table}[]
    \centering
    \begin{tabular}{l c c c c}
    \hline
      \multirow{2}{*}{Passband}  & \multicolumn{2}{c}{Solar-like star} & \multicolumn{2}{c}{A star} \\ 
       & $u_1$ & $u_2$ & $u_1$ & $u_2$ \\
      \hline
       V (550\,nm) &  0.517 &  0.229 &   0.276  & 0.388 \\
       R (650\,nm)  &  0.395 &   0.279 &  0.216 &  0.359 \\
       J (1200\,nm) &  0.149 &  0.334 &  0.064 &  0.300 \\
       H (1650\,nm) &  0.058 &  0.375 &  0.031 &  0.276 \\
       K (1650\,nm) &  0.041 &  0.327 &  0.015 &  0.236 \\
       L (3450\,nm) &  0.019 &  0.240 &  0.006 &  0.132 \\
       \hline
    \end{tabular}
    \caption{Limb-darkening coefficients for the two stars considered in the light curve simulations, in the passbands where these calculations were carried out.}
    \label{tab:ldc}
\end{table}

\begin{figure*}[ht!]
    \centering
    \hbox{\includegraphics[width=0.5\columnwidth]{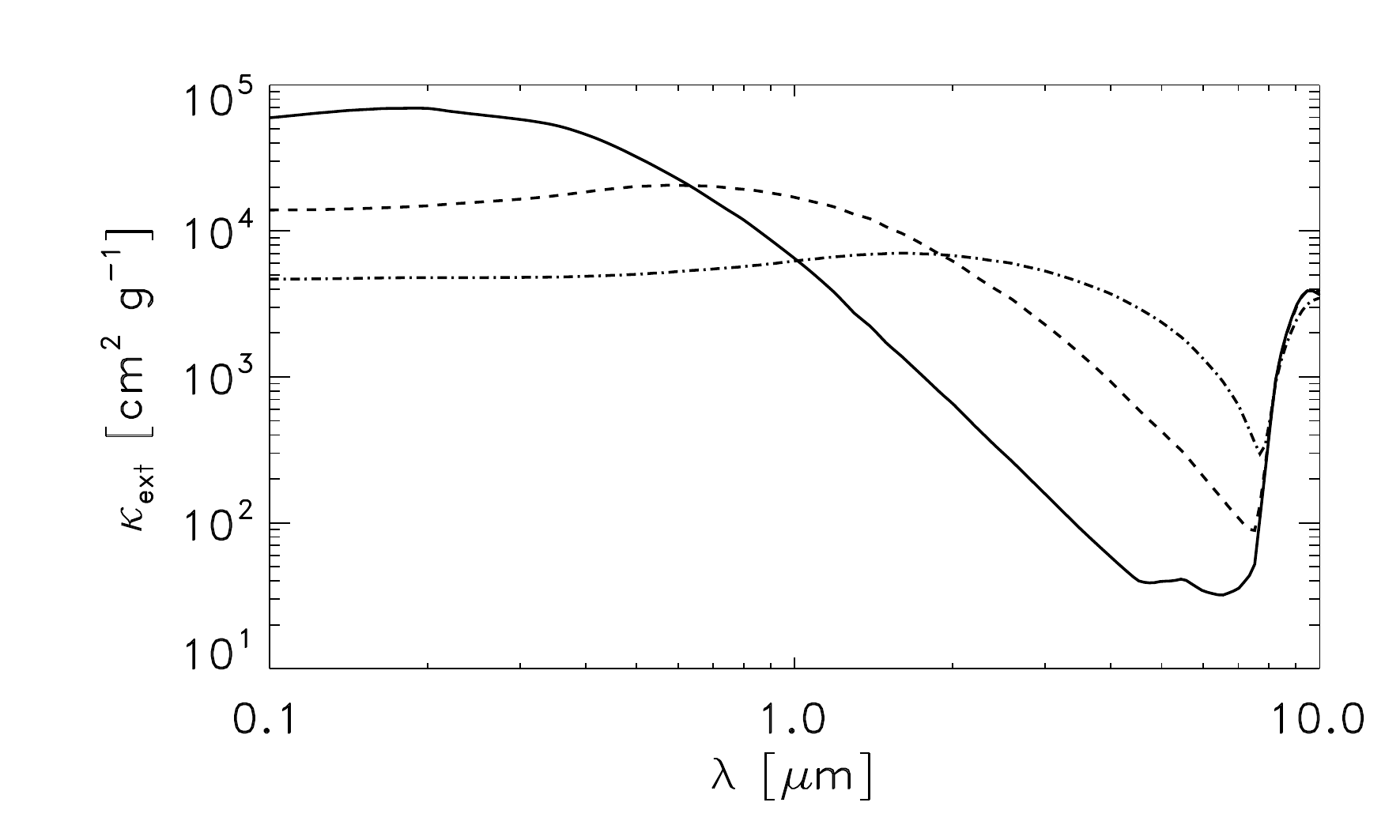}
    \includegraphics[width=0.5\columnwidth]{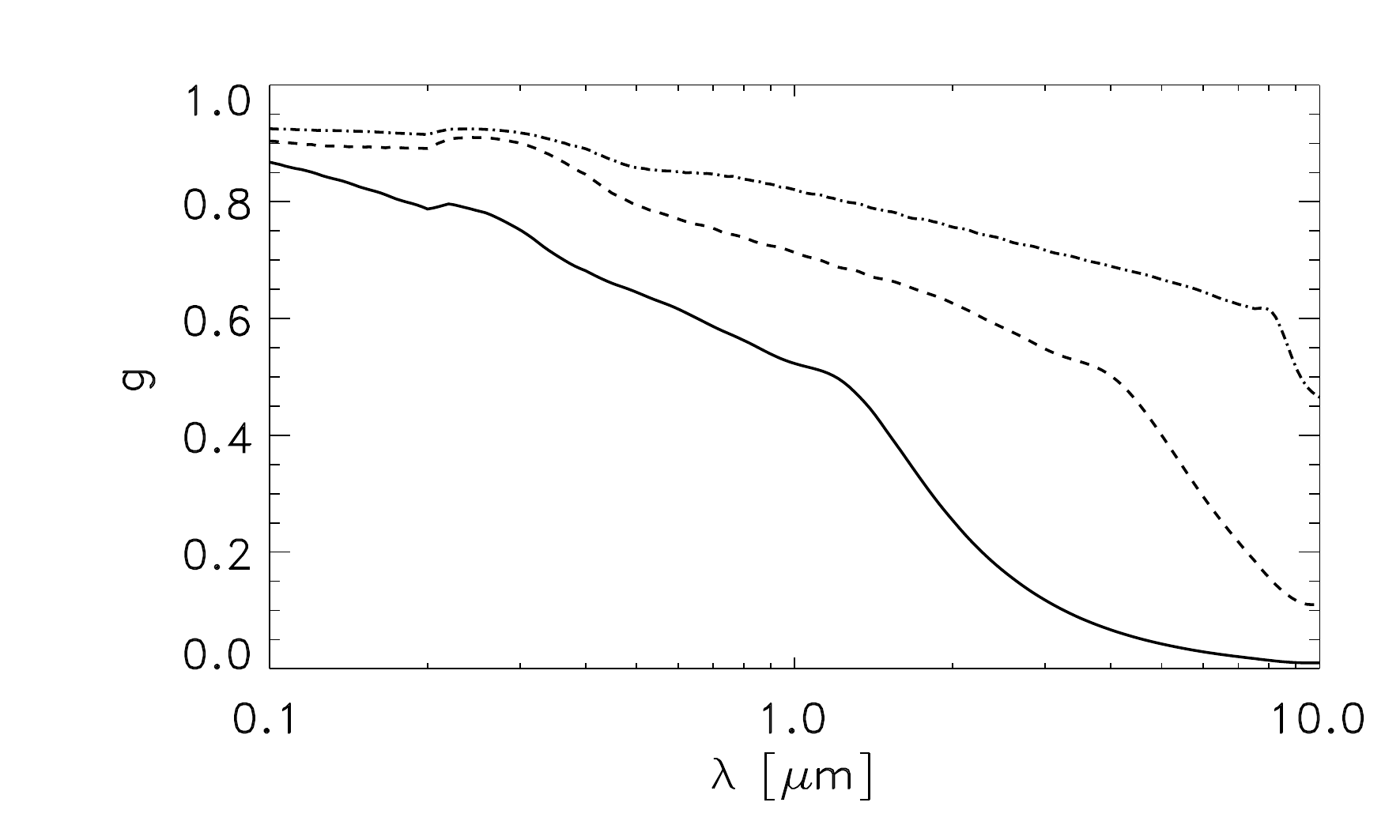}}
    \caption{Left panel: Extinction coefficient as a function of wavelength for pyroxene grains with 70\% Mg-content \citep{Dorschner1995,Henning1996}. Right panel: $g$ scattering parameter of the Henyey-Greenstein function as a function of wavelength, for different grain sizes, for the same material as in the left panel.  The curves on both panels correspond to grain sizes of 200\,nm (solid), 650\,nm (dashed) and 2\,$\mu$m (dash-dotted). }
    \label{fig:kappa}
\end{figure*}

Extinction (defined as absorption plus scattering) efficiency of the dust grains depends on the size and material of a grain, as well as the phase angle. As the same amount (mass) of dust may consist of a larger number of smaller grains or smaller number of larger grains, 
in a random spatial distribution smaller grains cover a larger area, with an area increase proportional to the relative grain size, leading to an overall increased extinction cross section. However, this is complicated by the grain size dependence of the extinction efficiency that depends on the dielectric properties of the grains. This is demonstrated in Fig.~\ref{fig:kappa} (left panel) where we show the extinction coefficient for pyroxene grains, one of the materials used in our simulations, for different grain sizes. 

Scattering by dust grains is typically strongly phase angle dependent in the visible -- near-infrared wavelength range \citep[see e.g.][]{Gordon2004}. The phase angle dependence in our simulations is characterized by the $g$ scattering parameter of the Henyey-Greenstein scattering function -- $g\,=\,0$ corresponds to isotropic scattering, $g\,=\,-1$ dominant backward scattering, and $g\,=\,1$ dominant forward scattering. An example is presented in Fig.~\ref{fig:kappa} (right panel) for pyroxene grains, which show strong forward scattering for visible light, and moderate forward scattering for the near-infrared. Due to our modeling geometry -- the dust structure is between us and the star, seen under a phase angle of $\phi$\,$\approx$\,180\degr\, -- our 'observed' light is dominated by forward scattered radiation. 
\begin{figure}[!h]
    \centering
    \includegraphics[width = \textwidth]{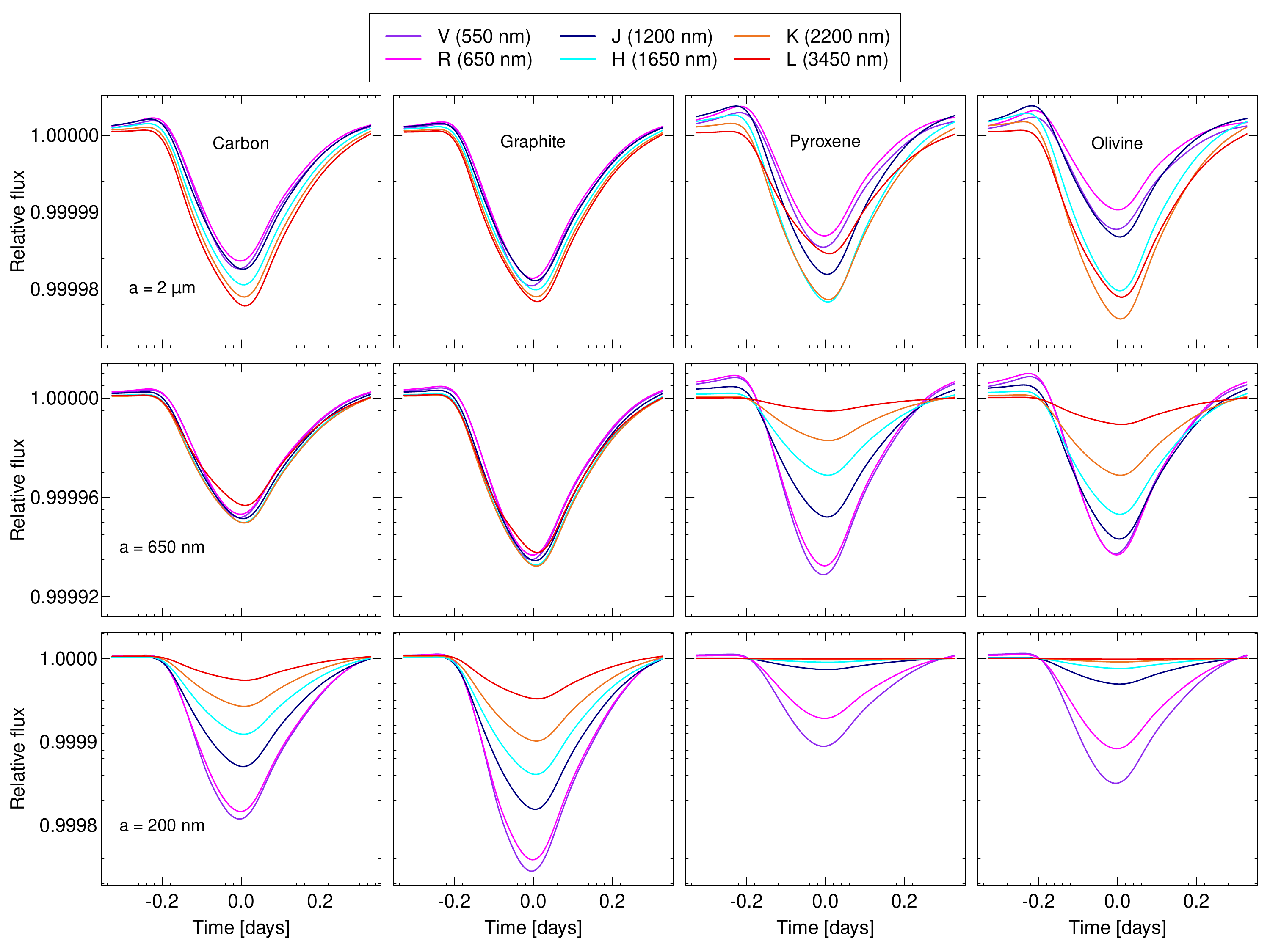}
    \caption{Synthetic, monochromatic light curves of a comet with narrow tail, plotted for the three selected grain size distributions, labelled as $a \sim 2$~$\mu$m (top row), $650$~nm (middle row) and $200$~nm (bottom row), for an A star. Columns denote the materials used in the simulations: carbon (left), graphite (second from the left), pyroxene (second from the right), and olivine (right). Note the change in the scale of the y axis from row to row.}
    \label{fig:73p_astar}
\end{figure}
Thermal emission of the dust is also included in the radiative transfer calculations. The dust temperature is also dependent on the composition and grain size \citep[see e.g.][]{Henning1996}, and due to the proximity of the dust structure to the central star ($r_h$\,=\,0.1\,au) relatively high $T_d$\,$\approx$\,1000\,K dust temperatures are expected. However, as in our simulations the distance to the star remains constant during the transit and the optical depths are low ($\tau$\,$\ll$\,1), the thermal emission contribution to the total brightness at a specific wavelength is constant before, during, and after the transit, meaning that it has no effect on the differential light curves.

The final ``observable'' light curve is the sum of the extinction and (forward--)scattering light curves. 
The calculated monochromatic light curves represent six common photometric bands: V (550\,nm), R (650\,nm), J (1.2\,$\mu$m), H (1.65\,$\mu$m), K (2.2\,$\mu$m) and L (3.45\,$\mu$m). We extrapolated the limb darkening coefficients for the L band (Table \ref{tab:ldc}), which was not included in \cite{1998A&A...335..647C}. These passbands were selected to represent the passband of \verb|PLATO| \citep{2014ExA....38..249R, 2024A&A...681A..18J} and to sample Ariel spectrographs \citep{2018ExA....46..135T,2021arXiv210404824T, 2022ExA....53..607S}, while avoiding the computationally expensive approach of simulating entire wavelength ranges of specific instruments.

\subsection{Density distributions \label{sect:dd}}

\begin{figure}[!h]
    \centering
    \includegraphics[width = \textwidth]{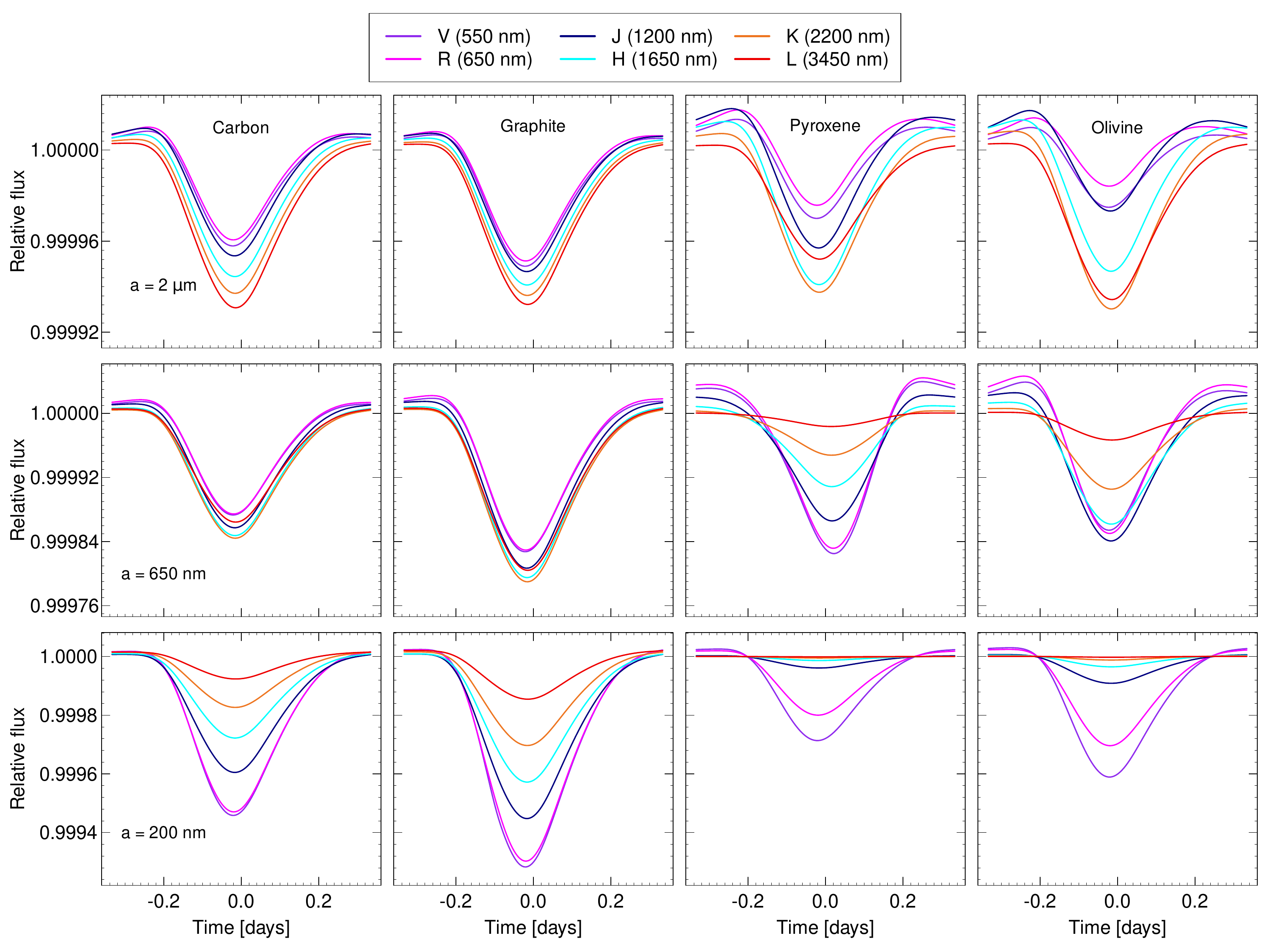}
    \caption{Same as Fig. \ref{fig:73p_astar} but for the comet model with a fan-like tail.}
    \label{fig:29p_astar}
\end{figure}

The input to our radiative transfer code is the spatial distribution of grains in the coma and possibly the tail in the comet. In the current stage, we can calculate the radiative transfer in a homogeneous environment built up by grains of a selected material with the same distribution of grain sizes everywhere in the coma. 

First, our idea was to put actual images of selected Solar System comets directly in transit in front of an arbitrary star. However, this concept had incurable drawbacks and therefore had to be omitted. The most important issues were: i) The comets are seen in a 2D projection on the sky, while the radiative transfer has to be calculated in a 3D environment -- including a third dimension in the distribution would have been mostly speculative rather than plausible; and ii) The scattering physics in a comet is complex. We see the actual comets because of the reflection properties of the dust at some specific illumination geometry, while we see the transiting exocomets because of the combination of absorption at a different, backlit geometry, together with components belonging to forward scattering. Without exact knowledge of material properties (spatially changing dust distributions, etc.) within the comet, the visible image cannot be plausibly converted into light scattering by that comet in transit.

Instead of using a combination of 2D images and additional more-or-less unfounded assumptions, we have chosen to base our model calculations entirely on mathematical foundations. This approach aligns well with the strength of our code, allowing us to calculate 3D distributions with homogeneous material properties. However, to ensure that our models are connected to the appearance of real comets, we used the visual appearance of observed Solar System comets as an inspirational source in the model design. We also aimed for a simple mathematical formalism and ultimately defined 3D distributions that accurately capture the impression of the chosen comets in their symmetry relationships and intensity distributions.

Our model parameters are therefore the material properties (which material, distribution of grain size) and the spatial density distribution. The actual cometary comae are known to follow a power-law light distribution that belongs to a spatial matter distribution of $r^{-2}$ in the case of homogeneous and isotropic matter production \citep{1984AJ.....89..579A}. This distribution has infinite extension and infinite mass -- which does not occur in practice, of course, due to dust dynamics and dust fragmentation in a radiation environment. Describing this cutoff is difficult if not impossible analytically, and in the case of real comets, dust dynamics is calculated with numerical methods \citep{2004come.book..565F}. In this paper, to avoid the issues of physical interpretation, we postulate that the truncating function is a Gaussian distribution with finite extension. Although not fitting to any comet, this truncated model describes a light distribution that resembles most cometary comae.


For the demonstration presented in this paper, we selected three models with different weights and symmetry of the tail. Instead of the physical model of cometary activity presented in \cite{1999A&A...343..916L} which conveniently recreates the shape of a classical comet as explained by the so-called ``fountain model'' \citep[e.g.][]{1957Tell....9...92A, 1968P&SS...16.1221W, 1981A&A....98...45W}, we construct dust density distributions that resemble well-known comets from the Solar System. As a result, we can also study the way the shape of an exocomet might influence it's transit light curve.

\paragraph{A comet with a narrow tail}

We regard one of the fragments of Comet 73P/Schwassmann-Wachmann as the baseline for our simulations. In this case, the comet is assumed to have a central coma and a long tail (Fig. \ref{fig:distributions}, left panels) to provide an appearance similar to that obtained by \citet{Reach2009}. We assumed a three dimensional density distribution in the form of
\begin{equation}
 \rho = \Bigg\{
 \begin{array}{lr}
    w_1 \rho_0 e^{-r^2/\sigma_1^2}/r^2 + (1-w_1) \rho_0 e^{-(x^2+z^2)/\sigma_2^2} e^{-y^2/\sigma_f^2} & : y \ge 0 \\
    w_1 \rho_0 e^{-r^2/\sigma_1^2}/r^2 + (1-w_1) \rho_0 e^{-(x^2+z^2)/\sigma_2^2} e^{-|y|/\sigma_t} & : y < 0 \\
  \end{array}
\label{eq:73p}
\end{equation}
to represent such a comet, where $r^2 = x^2 + y^2 + z^2$. We set $w_1\,=\,0.2$, $\sigma_1\,=\,0.0015$\,au, $\sigma_2\,=\,0.0025$\,au, $\sigma_f\,=\,0.0025$\,au and $\sigma_t\,=\,0.010$\,au. This dust distribution also outlines the limitations of our modeling, and the deviation from \citep{1999A&A...343..916L, 1999A&AS..140...15L, 2019A&A...625L..13Z}, since the tail of the cometary shape is truncated and we neglect the change in the distance between the star and each particle of the cometary tail.

\paragraph{A comet with a wide, fan-like tail} 

Here, a central dense coma is surrounded by a spherical structure that is extended in one direction (defined in the $+y$ direction in the simulation volume), but fades quickly in the opposite direction as in the case of Comet 29P/Schwassmann-Wachmann \citep{2008A&A...485..599T}. In this case, we assumed the following densitry distribution:
\begin{equation}
 \rho = \Bigg\{
 \begin{array}{lr}
    w_1 \rho_0 r^{-2} e^{-r^2/\sigma_1^2} + (1-w_1) \rho_0 e^{-r^2/\sigma_2^2} & : y \ge 0 \\
    w_1 \rho_0 r^{-2} e^{-r^2/\sigma_1^2} + (1-w_1) \rho_0 e^{-(x^2+z^2)/\sigma_2^2} e^{-y^2/\sigma_t^2}& : y < 0 \\
  \end{array}
\label{eq:29p}
\end{equation}
The relative weights of the central and extended components are characterized by the weight factor $w_1$, the core and extended structures are described by two Gaussians with width parameters $\sigma_1$ and $\sigma_2$, and the fading tail is also described with a Gaussian with the width parameter $\sigma_t$ in the ---- $y$ direction. We set $w_h\,=\,0.5$, $\sigma_1\,=\,0.003$\,au, $\sigma_2\,=\,0.012$\,au and $\sigma_t\,=\,0.006$\,au in our calculations. The resultant distribution is shown in Fig. \ref{fig:distributions}, in the middle panels.

\paragraph{A comet without tail} 
This density distribution is assumed to be fully spherically symmetric, that is, the density depends only on the radial distance $r$ from the center of the comet. This density distribution was selected to resemble Comet 17P/Holmes during its 2007 outburst \citep[e.g.][]{2008ApJ...677L..63M, 2008ApJ...684L..55M, 2008LPI....39.1627T}. The distribution is constructed as the sum of two 3D Gaussian profiles, characterized by the width parameters $\sigma_1$ and $\sigma_2$. The relative weights of the two Gaussians are determined by the weight factor $w_h$. The central density $\rho_0$ is set for each simulation as described below. The spatial distribution of the density is determined as follows: 

\begin{equation}
    \rho = w_h \rho_0 e^{-r^2/\sigma_1^2} + (1-w_h) \rho_0 e^{-r^2/\sigma_2^2}
    \label{eq:holmes}
\end{equation}
We applied $w_h$\,=\,0.7, $\sigma_1$\,=\,0.002\,au and $\sigma_2$\,=\,0.008\,au.

To set up the dust distributions, our requirement was that the maximum optical depth ($\tau$) in the simulation with the a\,=\,2\,$\mu$m grain size is $\tau_R$\,=\,10$^{-4}$ in the R band (650\,nm) for the solar-like star, corresponding to the nominal maximum absorption depth\footnote{We regard $\tau$ as the measure of the extinction component \textit{only}, and AD as the measure of the combination of \textit{both} light curve components (extinction and forward scattering). Thus, in a real observation scenario, only AD can be obtained.} (AD) identified by \cite{2022NatSR..12.5855L} in the  $\beta$ Pictoris system. This specifies a total dust mass that is then kept constant for the other grain sizes (a\,=\,200 and 650\,nm). These dust masses are listed in Table~\ref{table:simmass} below. In the following, we refer to the three dust distributions by the name of the comet that they are meant to resemble.

\begin{table}[ht!]
    \centering
    \begin{tabular}{lcccc}
 \hline
          Comet name & carbon [$10^{11}$\,kg]	& graphite [$10^{11}$\,kg] & pyroxene [$10^{11}$\,kg] & olivine	[$10^{11}$\,kg]\\
          \hline
           No tail & 2.96 & 4.14e+11 & 5.49e+11 & 6.98e+11 \\
Fan-like tail & 4.65 & 7.48 & 9.93 & 1.26 \\
Narrow tail & 0.84 & 1.35 & 1.80 & 2.29 \\
        \hline
    \end{tabular}
    \caption{Total dust masses of the four materials used in the simulations, at the distance of $r_h$\,=\,0.1\,au. }
    \label{table:simmass}
\end{table}


\section{Results} \label{sec:res}

\begin{figure}
    \centering
    \includegraphics[width = \textwidth]{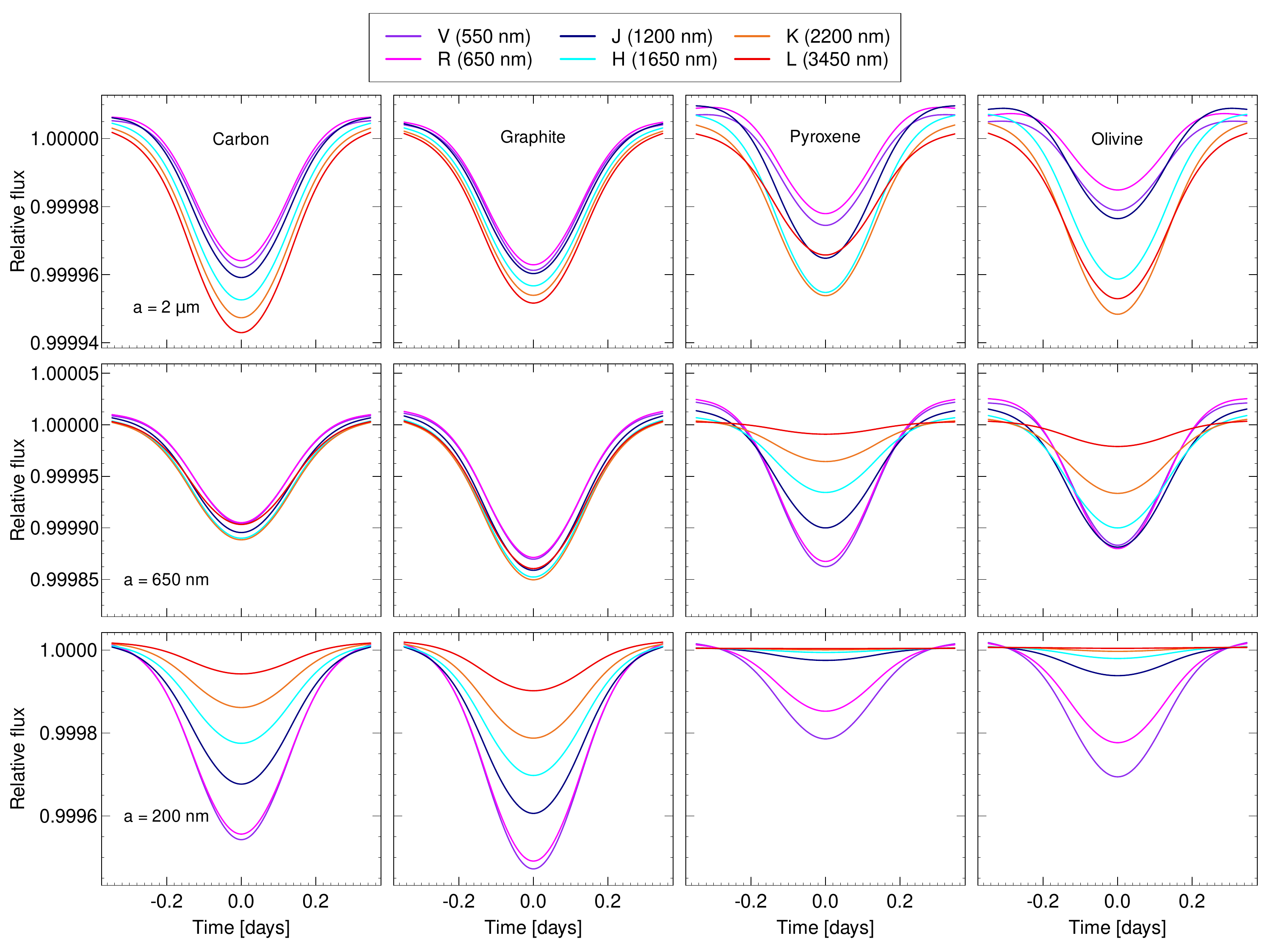}
    \caption{Same as Fig. \ref{fig:73p_astar} but for the comet model without a tail.}
    \label{fig:17p_astar}
\end{figure}

Simulated light curves were calculated for each of the three model distributions in transit, assuming carbon, graphite, pyroxene, and olivine dust, and with a characteristic grain size in the coma/tail of 2~$\mu$m, 650~nm and 200~nm. Transit light curves were calculated in the 6 predefined wavelengths (resulting in 12 scenarios for comet each model, at every wavelength), yielding a 6-color photometric time series in each simulation. Template light curves are plotted in Figs. \ref{fig:73p_astar} -- \ref{fig:17p_astar} for a star with spectral type A, and in \ref{fig:73p_v1}--\ref{fig:17p_v1} for a Solar-like star.

\subsection{Light curve morphology}

Figures \ref{fig:73p_astar} and \ref{fig:73p_v1} show the exocomet with a prominent thin tail, which is a morphologically somewhat similar distribution to what is expected from the fountain model \citep[e.g.][]{1957Tell....9...92A, 1968P&SS...16.1221W, 1981A&A....98...45W}, leading to sharkfin-like transit light curves. Indeed, on a qualitative comparison, the light curves in Figs. \ref{fig:73p_astar} and \ref{fig:73p_v1}  closely resemble those seen in \cite{1999A&A...343..916L} and \cite{1999A&AS..140...15L}. Our simulations also bear close resemblance to observed transit features that are attributed to exocometary activity, such as in \cite{2018MNRAS.474.1453R, 2019A&A...625L..13Z, 2022NatSR..12.5855L}, or a disintegrating exoplanet \citep{2012ApJ...752....1R,2012A&A...545L...5B}. The light curves seen in Fig. \ref{fig:73p_v1} and especially in Fig. \ref{fig:73p_astar} readily reproduce the peak observable before the transit itself \citep{2018MNRAS.474.1453R, 2019A&A...625L..13Z}, which is a consequence of the forward--scattered light.

Figs. \ref{fig:73p_astar} -- \ref{fig:17p_astar} and \ref{fig:73p_v1}--\ref{fig:17p_v1} unanimously exhibit a common strong dependence of $\tau$ (and, as a consequence, AD) on the dust grain size, where smaller grain sizes generally yield deeper transits. This is especially true for carbon and graphite. This is a natural consequence of our calibration process, since all distributions of individual comets have the same total mass (Table \ref{table:simmass}). Figs. \ref{fig:73p_ads}, \ref{fig:29p_ads}, and \ref{fig:17p_ads} represent the measured AD for the three comet models considered. The well-known wavelength dependence of the scattering cross section (that is itself $\propto a^2$) means that while the light curves of a given material for a given $a$ are topologically similar to each other, in the regime of small dust grains ($a \sim 200$~nm), the absorption depth varies considerably throughout the explored passbands. At the $2$\,$\mu$m dust size range, there are only comparatively minor differences in absorption depths at different wavelengths. Slightly larger differences in AD can be observed at the $650$\,nm grain sizes, where (unlike in $200$\,nm regime) AD increases towards the red end of the wavelength range (Figs. \ref{fig:73p_ads} -- \ref{fig:17p_ads}).

\begin{figure}
    \centering
    \includegraphics[width = .32\textwidth]{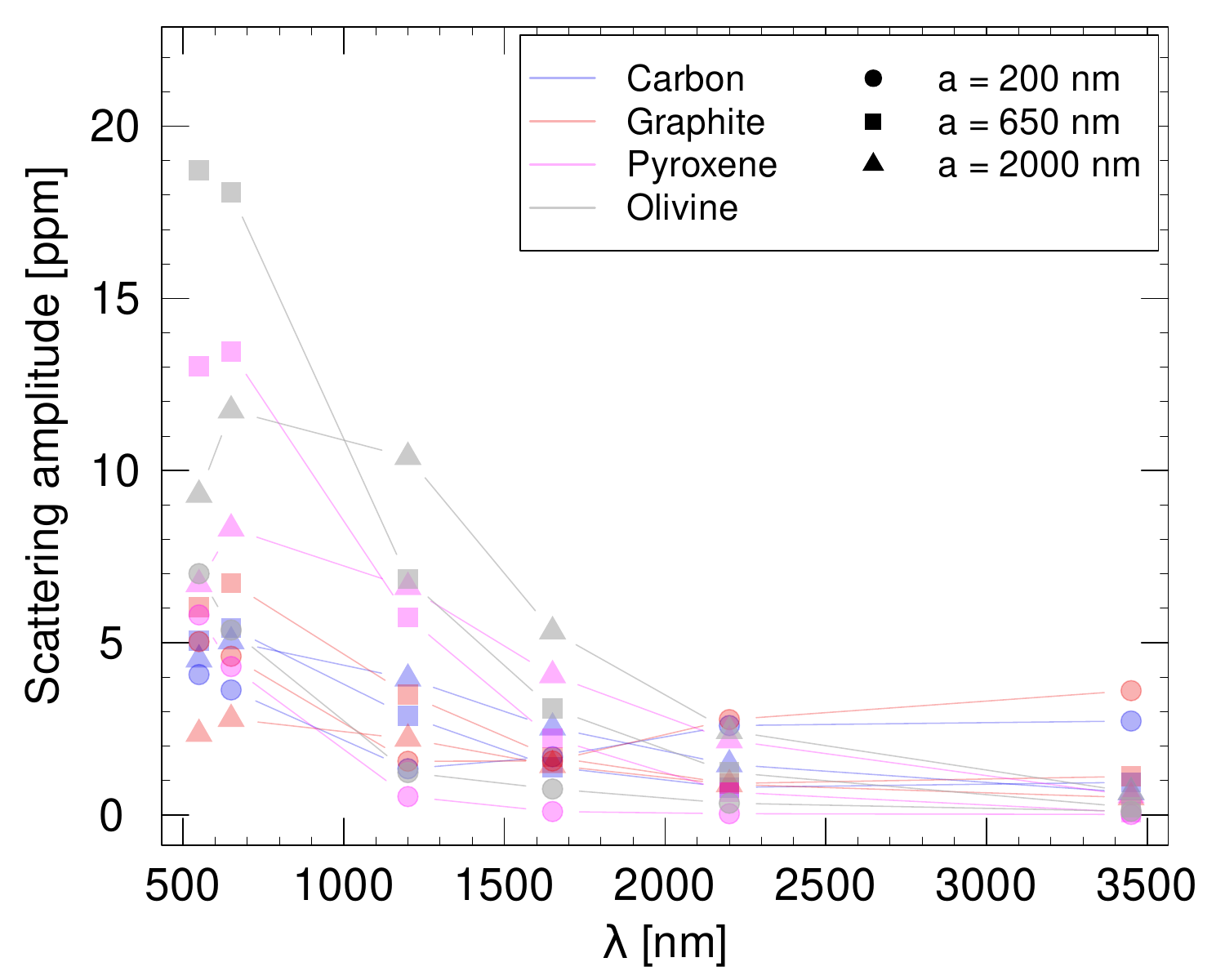}
    \includegraphics[width = .32\textwidth]{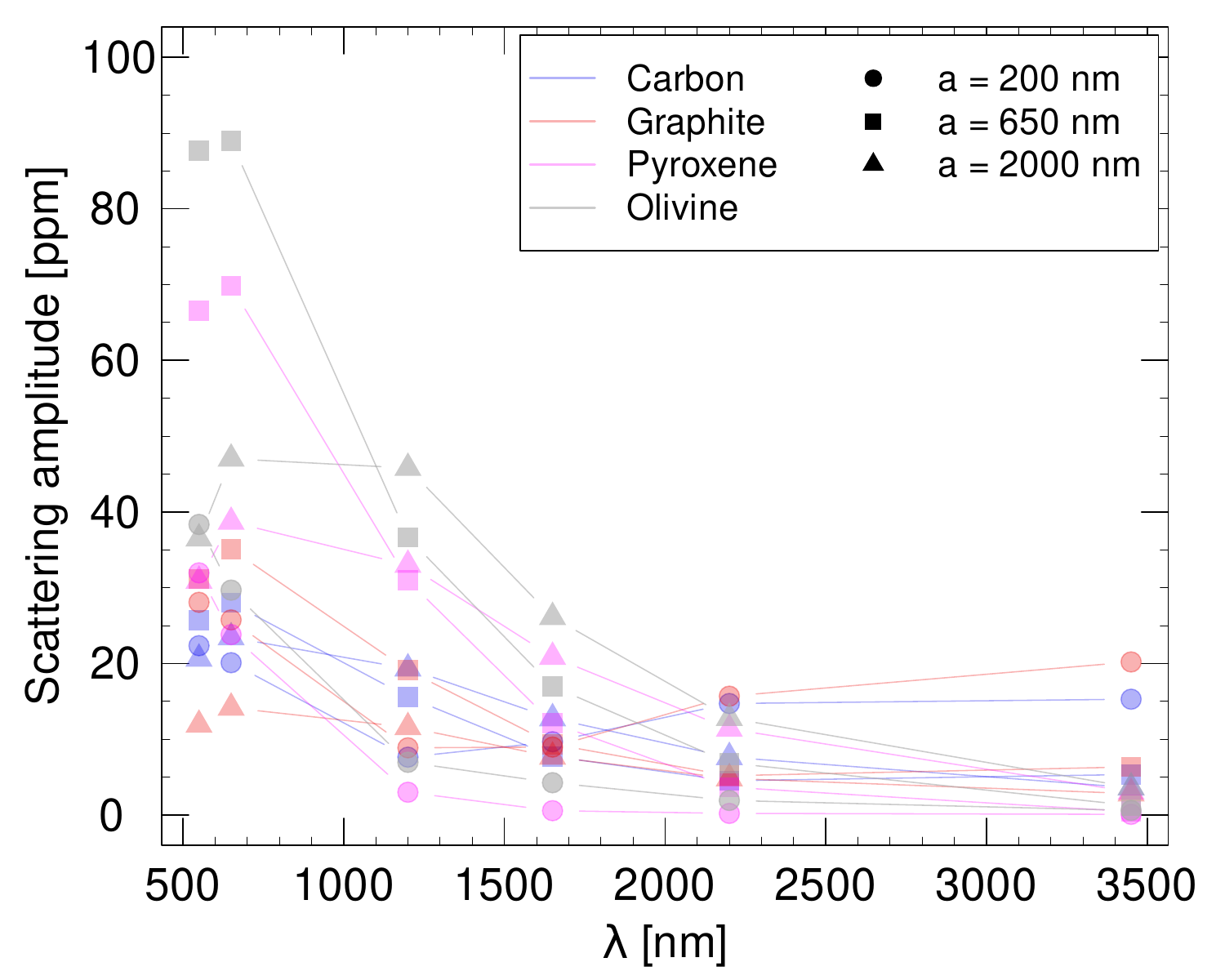}
    \includegraphics[width = .32\textwidth]{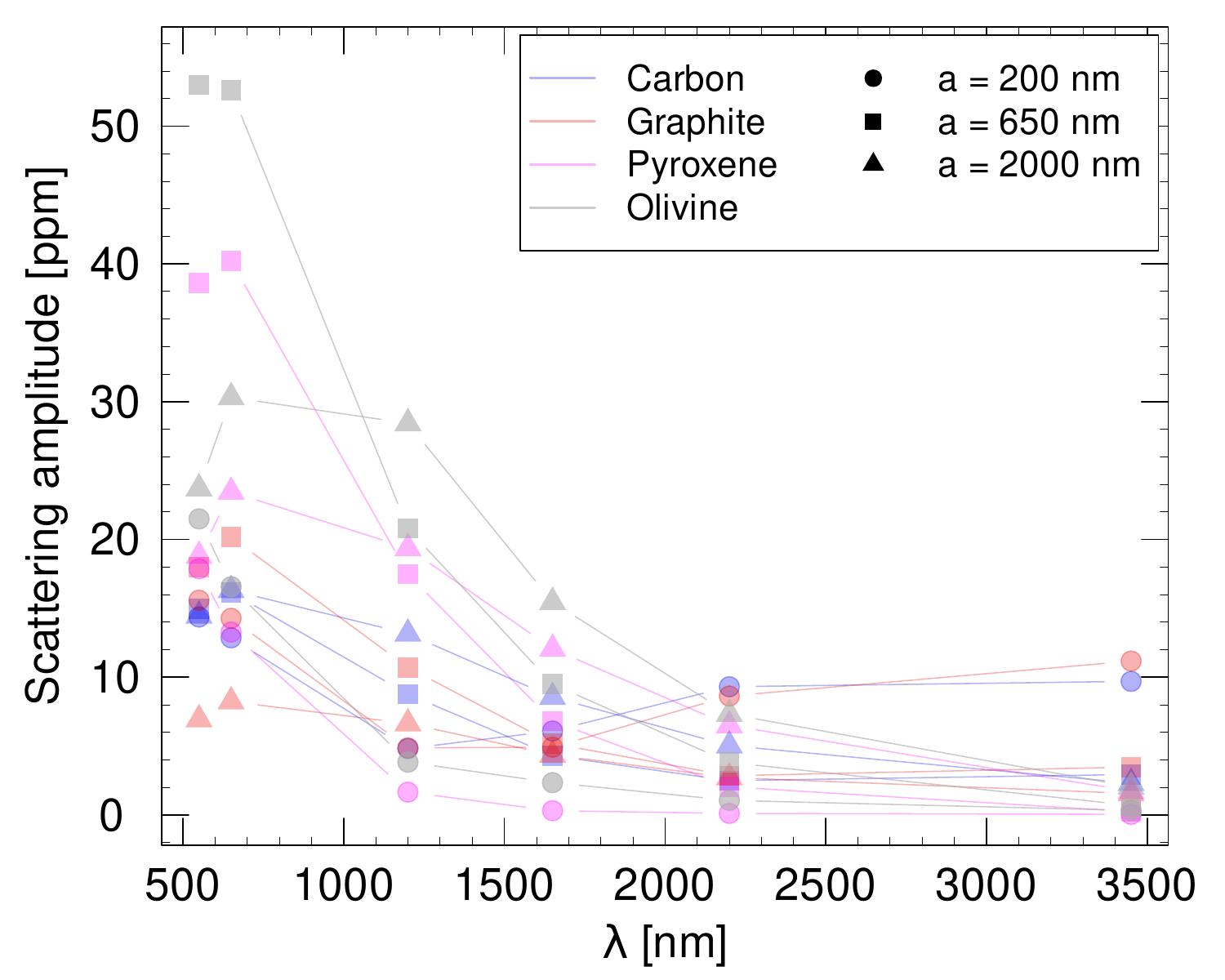}
    \caption{Amplitudes of the forward-scattered light for the three dust distributions shown on Fig. \ref{fig:distributions} (narrow-tailed comet -- left, fan-like-tailed comet -- middle, tailles comet -- right) for the A star. Note the change in the scale of the y axis. Straight lines are shown for visualization purposes only.}
    \label{fig:amplitudes}
\end{figure}

In all three cases, we see excess light before transit ingress. This is the contribution of forward scattering, the light coming from the dust grains, which are seen close to the edge of the star but are still not in transit. Because forward scattering is efficient at very small deflection angles (Fig. \ref{fig:kappa}), this contribution appears just before the transit (and is well known in the case of transits of evaporating/disintegrating planets as well \citep[e.g.][]{2012ApJ...752....1R,2018A&A...611A..63G}. The amplitude of the forward-scattered light also shows a clear dependence on both the material composition, the grain size, and the wavelength, dust distribution, and stellar spectral type (Fig. \ref{fig:amplitudes}), with the highest amplitude (on the order of $\approx 90$\,ppm) occurring in the V and R bands, for olivine, in the case of the wide-tailed comet model. Fig. \ref{fig:amplitudes} also clearly shows that carbonaceous materials are less efficient at forward scattering, implying that the exocomet candidates observed by \citep{2019A&A...625L..13Z} are not primarily made of these materials. 
Infrared wavelengths yield lower amplitudes for this component of the light curve.  In this regime we observe a complete turnaround of the amplitudes as the $a = 200$\,nm grain sized dust shows the highest amplitudes for the forward-scattered light, for the carbonaceous material.

The amplitude is also influenced by the shape of the comet itself, this manifests in the fact that the wide-tailed comet (Fig. \ref{fig:distributions}) yields the highest amplitudes for this light curve component. Both comet models that own a tail component produce asymmetrical scattered light components, unlike the spherically symmetric coma alone.

A prominent difference between extinction and forward scattering is that the latter depends on the direct illumination of the dust. Therefore, the amplitude of the forward scattering components are scaled to arbitrary star--comet distances by $\left(r_h/1 \text{au} \right)^2$ as a first-order approximation (where $r_h$ is the actual star--comet distance). In theory, the transit duration can be used to estimate $r_h$ from a real exocomet transit observation \citep[e.g.][]{2023A&A...671A.127B}, which could then be used to put further investigate the cometary composition through the forward-scattered light. Because the extinction simply occurs due to blocking of a part of the line of sight, the extinction components (analogous to the transit depths) are independent of $d$ \citep[e.g.][]{2002ApJ...580L.171M}, and depend only on the total scattering cross section of the dust. In this way, proper modeling of the forward scattering and the extinction together can also be used to reveal the star-to-transient distance via the ratio of the extinction to forward scattering.

Figs. \ref{fig:29p_astar}--\ref{fig:17p_astar} and \ref{fig:29p_v1}--\ref{fig:17p_v1} show transient light curves for comet models with increasing circular symmetry, that is, a fan-like tail and no tail at all. 
The combined light curve of the single coma (Figs. \ref{fig:17p_astar} and \ref{fig:17p_v1}) also resembles a grazing transit of an exoplanet \citep[e.g.][]{2014A&A...568A.127M}. This is an important observation, because the shape of the transient itself is not a decisive diagnostics of the comet nature: exocomets (being similar to 17P/Holmes in outburst or other coma-dominated comets such as 1P/Encke for example) can also produce symmetrical light curve transients. On the other hand, possibly dust clumps (in a debris disk) unrelated to comets can also bear asymmetry that results in a more sharkfin-like transient feature. In these cases, we propose to observe the AD ratios (Figs. \ref{fig:73p_ads} -- \ref{fig:17p_ads}), or the ``colors'' directly during transit (\ref{fig:73p_solar_colors} -- \ref{fig:h2_astar_colors}). The wide range in AD that is observed especially for the smallest grain sizes represented here (Figs. \ref{fig:73p_ads} -- \ref{fig:17p_ads}) implies that a spectrophotometric observation can be used to extract information regarding the material composition of an exocometary transit. This is discussed below.

The light curves are similar to each other between the two stars (a `solar-like' and an A-type star) for all three given dust distributions. This is observed when comparing Figs. \ref{fig:73p_astar}, \ref{fig:29p_astar}, and \ref{fig:17p_astar} with \ref{fig:73p_v1}, \ref{fig:29p_v1}, and \ref{fig:17p_v1} , respectively. One key difference is the Absorption Depth, as the same dust distribution transiting in front of the A spectral type star considered here (that is $1.7\times$ larger than the Sun-like star) yield transits that are $\leq 2 \times$ shallower at each wavelength. In the cases of exoplanets, the planet is fully obscuring a portion of the stellar disk, leading to a direct way to estimate its size based on the transit depth \citep[e.g.][]{2003ApJ...585.1038S}, which is analogous to AD. For exocomets, no such straightforward method can exist to estimate the size based on the absorption depth, as it depends on a large number of factors, as shown above. We have also repeated the same calculations with a Solar-like host star, and we show the light curves in Appendix {\ref{sect:app_lc}}.

\subsection{Color index variations}

\begin{figure}
    \centering
  \includegraphics[width = 0.48\textwidth]{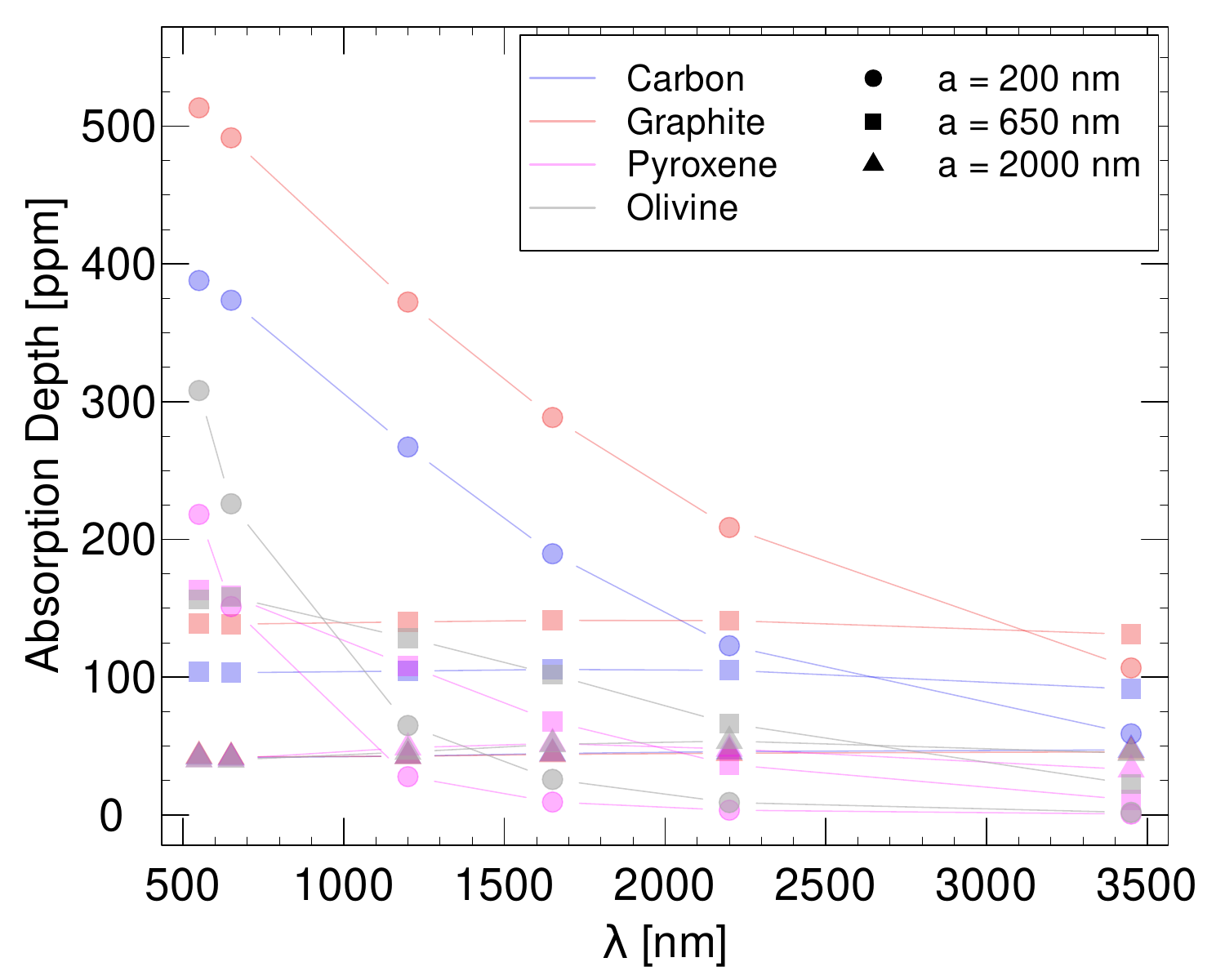}
  \includegraphics[width = 0.48\textwidth]{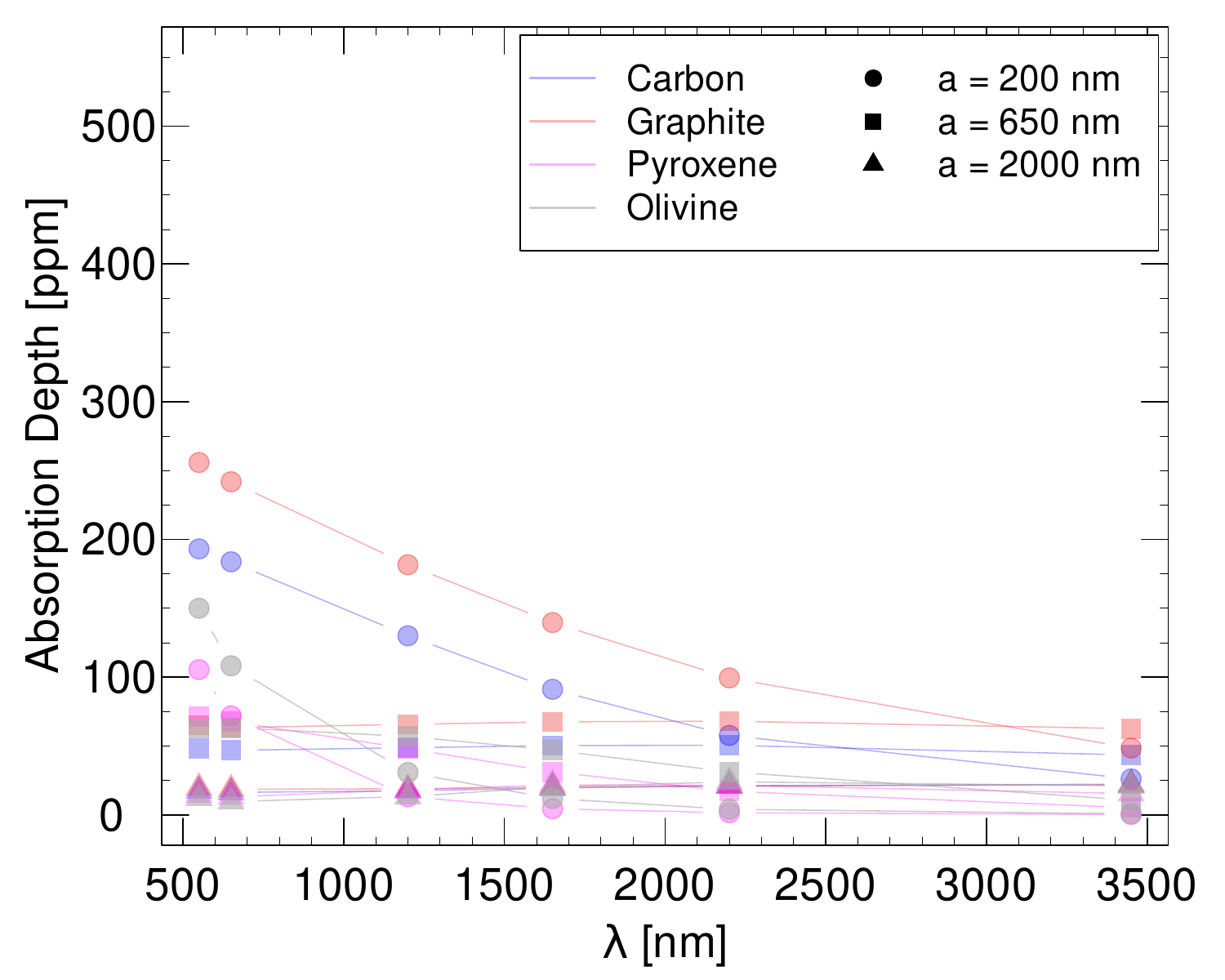}
    \caption{Absorption depth as a function of wavelength for the narrow-tailed comet (Fig. \ref{fig:distributions}, left) for the Solar-like star (left) and the A star (right). The four materials considered (carbon, graphite, pyroxene, and olivine) are represented by colors (blue, red, magenta, and grey, respectively), while full circles, squares, and triangles correspond to the different dust grain sizes ($200$\,nm, $650$\,nm, and $2$\,$\mu$m, respectively). Straight lines are shown for visualization purposes only.}
    \label{fig:73p_ads}
\end{figure}

\begin{figure}
    \centering
  \includegraphics[width = 0.48\textwidth]{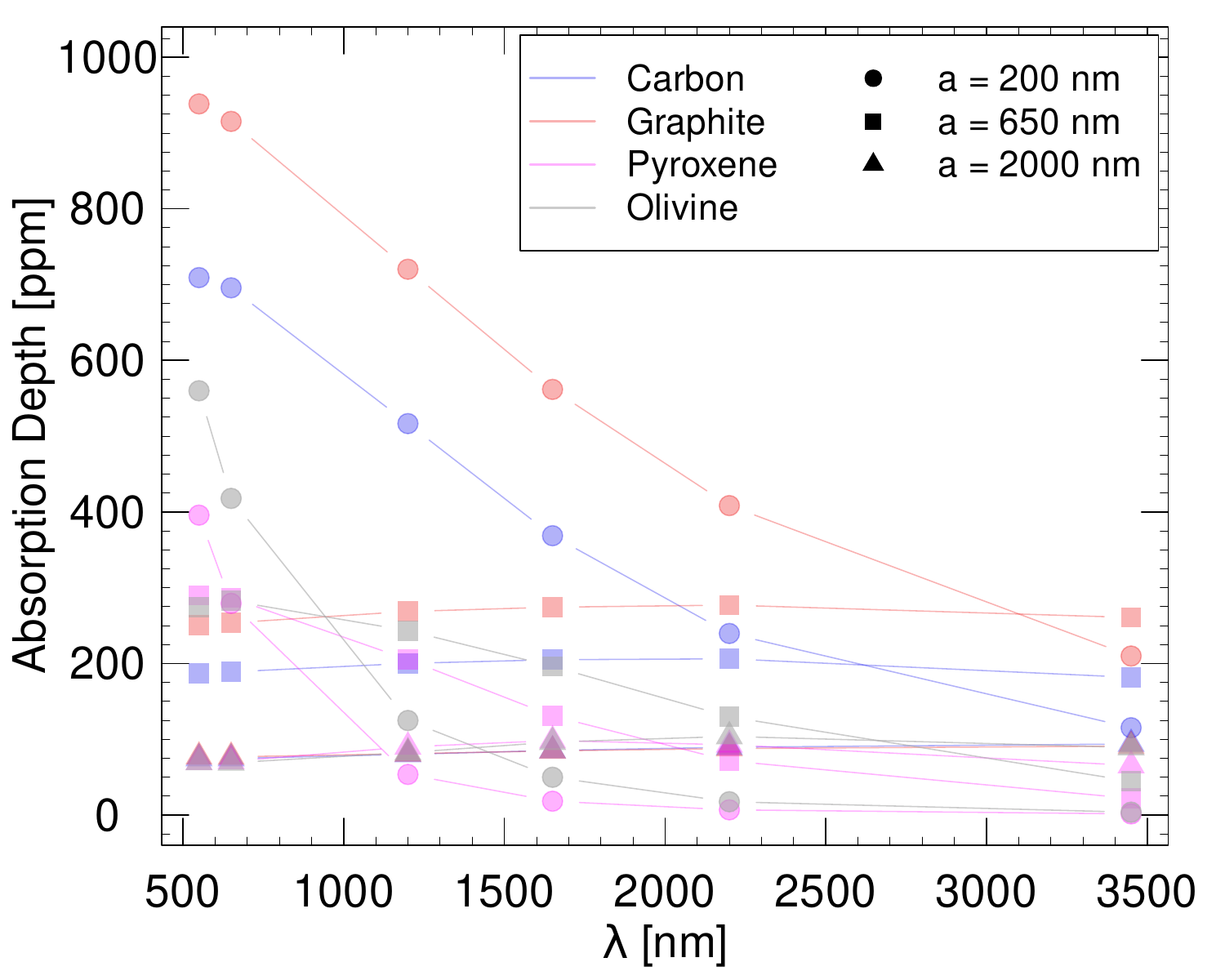}
  \includegraphics[width = 0.48\textwidth]{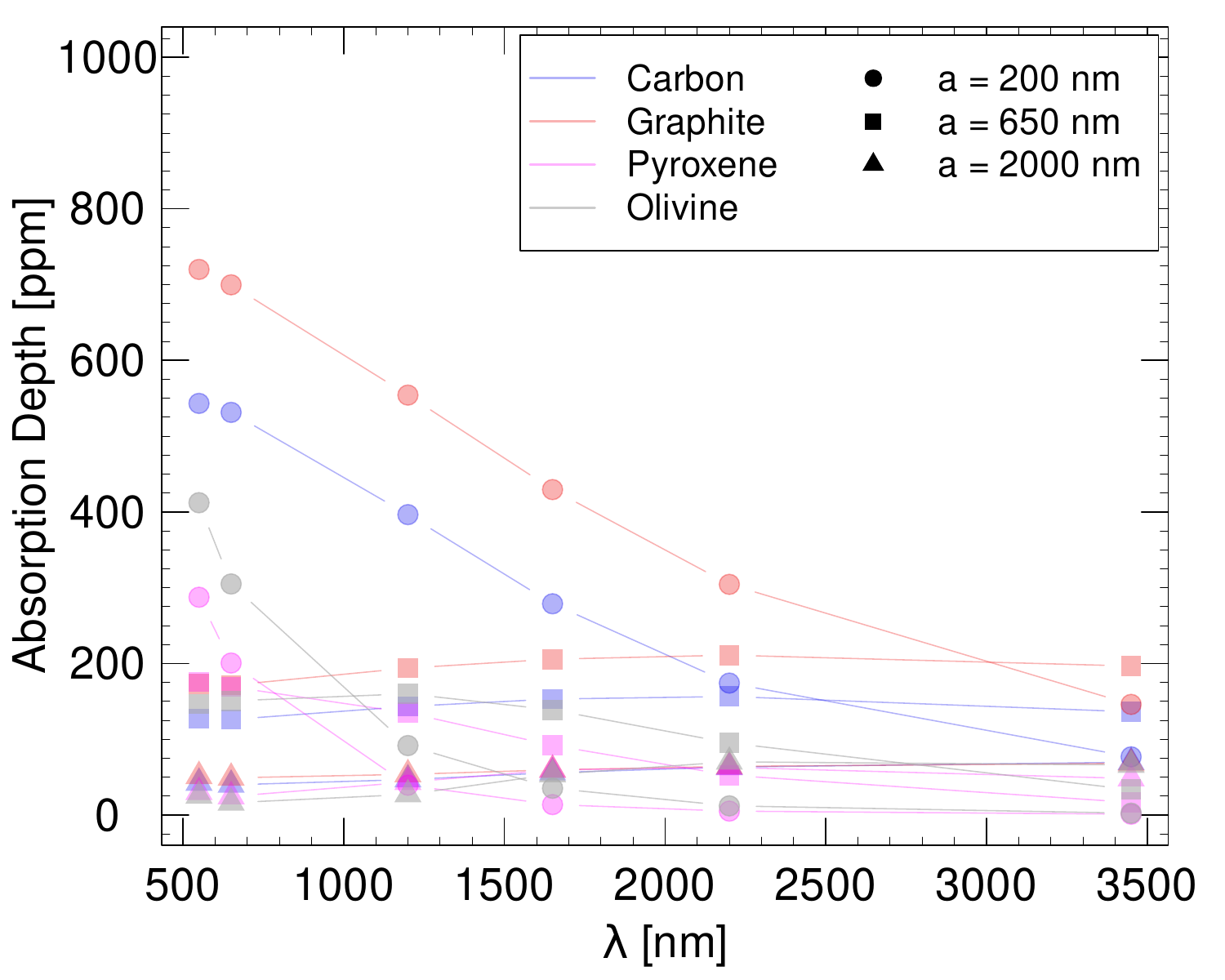}
    \caption{Absorption depth as a function of wavelength for the narrow-tailed comet (Fig. \ref{fig:distributions}, left) for the Solar-like star (left) and the A star (right). The four materials considered (carbon, graphite, pyroxene, and olivine) are represented by colors (blue, red, magenta, and grey, respectively), while full circles, squres, and triangles correspond to the different dust grain sizes ($200$\,nm, $650$\,nm, and $2$\,$\mu$m, respectively). Straight lines are shown for visualization purposes only.}
    \label{fig:29p_ads}
\end{figure}

When comparing Figures \ref{fig:73p_astar}--\ref{fig:17p_astar} and \ref{fig:73p_v1}--\ref{fig:17p_v1}, the dependence of AD on the properties of the material and the characteristic size of the particles is evident. This is further emphasized by the direct comparison of AD (Figs. \ref{fig:73p_ads}, \ref{fig:29p_ads}, and \ref{fig:29p_ads}), or the color variations (Figs. \ref{fig:73p_solar_colors} -- \ref{fig:h2_astar_colors}). The asymmetry of the dust distribution in combination with the (wavelength-dependent) limb darkening of the star yields $\approx 6$ time brighter or fainter portions of the light curve (Figs. \ref{fig:73p_solar_colors} and \ref{fig:73p_astar_colors}). This is a phenomenon previously known and extensively studied in the case of solar system comets \citep[e.g.][]{1986ApJ...310..937J,1997Icar..126..351K,2019MNRAS.485.4013L}. However, the available time series data for extrasolar comets are currently monochromatic, so the value of AD ratios and colors has not been particularly emphasized until now. 

\begin{figure}
    \centering
    \includegraphics[width = 0.48\textwidth]{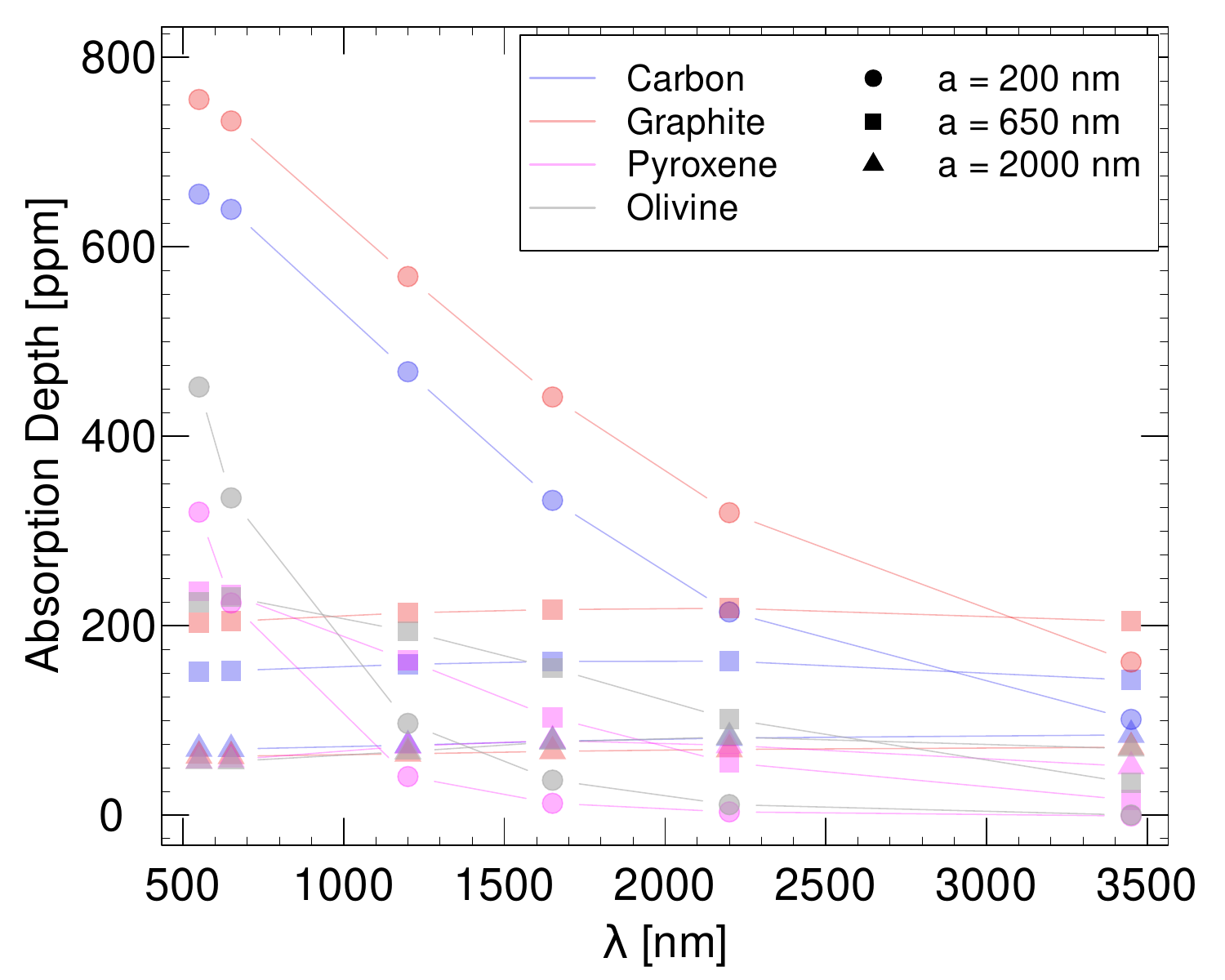}
    \includegraphics[width = 0.48\textwidth]{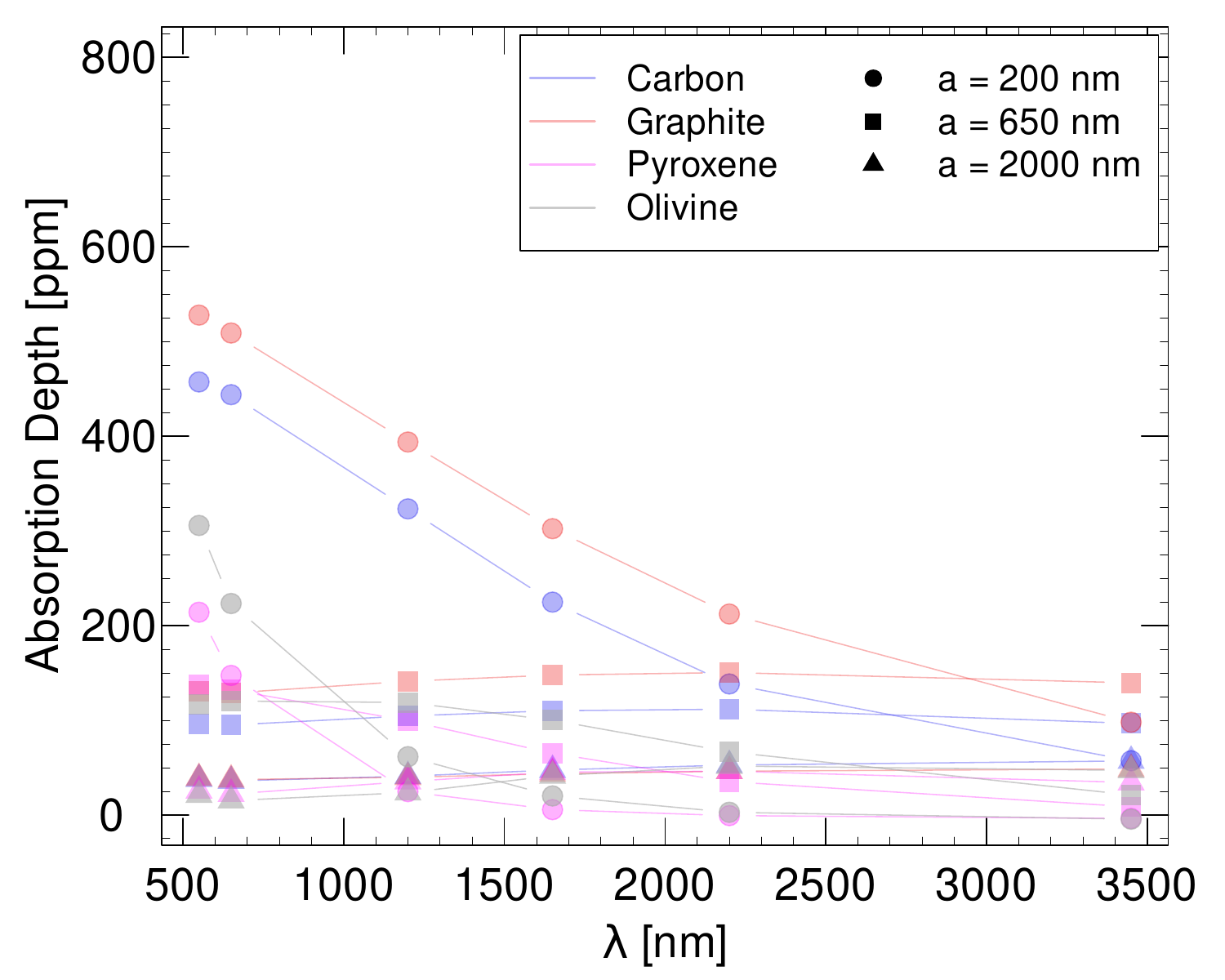}
    \caption{Absorption depth as a function of wavelength for comet without tail (Fig. \ref{fig:distributions}, right) for the Solar-like star (left) and the A star (right). The four materials considered (carbon, graphite, pyroxene, and olivine) are represented by colors (blue, red, magenta, and grey, respectively), while full circles, squres, and triangles correspond to the different dust grain sizes ($200$\,nm, $650$\,nm, and $2$\,$\mu$m, respectively). Straight lines are shown for visualization purposes only.}
    \label{fig:17p_ads}
\end{figure}

We can observe that the smallest dust grains ($\sim$ 200~nm size range) produce transit light curves of significantly varying depth with the central wavelength. This is a ``colorful'' dust in transmission. In the case of carbonaceous dust (carbon and grahite) with 200 nm size, the optical V and R signals are the deepest, and they are very similar to each other, the V-R color index remains neutral during the transit, while the NIR components exhibit shallower transit light curves, which are material independent. We can observe the transit even in the L band as well, unlike in the case of silicates. When dust has a size range of 2~$\mu$m, the deepest signal is observed in the L band, the optical transit light curves are the most shallow ones, but with much less color difference than in the case of small dust. In between, at around 650 nm dust size ranges, the transit curves follow a very similar shape with close-to-neutral colors (i.e. near-constant AD, or color index of 1, Figs. \ref{fig:73p_ads}--\ref{fig:17p_ads}, and \ref{fig:73p_solar_colors} -- \ref{fig:h2_astar_colors}). In summary, the same mass of carbonaceous dust in transmission is ``dark red'', ``less dark neutral'' and ``pale blue'' if the size range is $\sim 200$, $650$ and $2000$ nm, respectively.

In the case of silicates, almost no variation is observed in the K and L bands in the case of the smallest dust distribution examined. The optical signal can be quite deep ($\sim 400$--$6000$~ppm in Figure \ref{fig:29p_ads}, for 29P orbiting a solar-like star), but still less deep than in the case of carbonaceous materials. With increasing dust size, AD in the $NIR$ bands also increases. In the case of the 2$\mu$m dust size, where the size of the scattering particles is compatible with the wavelength, the deepest signal will be observed in the H and K bands for pyroxene and olivine, respectively.

The change in color index becomes truly significant when the photometric transients are detected simultaneously in multiple bands. While differences can be seen by comparing the light curves, e.g. in Figs. \ref{fig:73p_astar}
--\ref{fig:17p_astar}, we quantitatively show the color variation curve in Appendix (\ref{sect:app_colvar}).

Exocomets are typically found in dusty environments, and transients related to the transit of overdensities in the hot dust cloud can lead to similar, nonperiodic transients. A substantial distinction between the two dust overdensity structures lies in the chemical differences and the different size distributions that can be discerned by multi-band observations. Dust clouds evolved in a warm environment and represent processed material, unlike exocomet dust, which originated freshly from a previously frozen environment and then underwent fragmentation, possibly presenting dust that has a different composition and grain size distribution than the warm dusty environment. Consequently, we anticipate that the transients of dusts will be only accompanied by minor changes in AD ratios at most, while exocomets will exhibit characteristic AD ratio changes that will also be consistent with our implemented code. This color information can distinguish between the two transient scenarios.

\subsection{Transit timing variations}

\begin{figure}
    \centering
    \includegraphics[width = .49\textwidth]{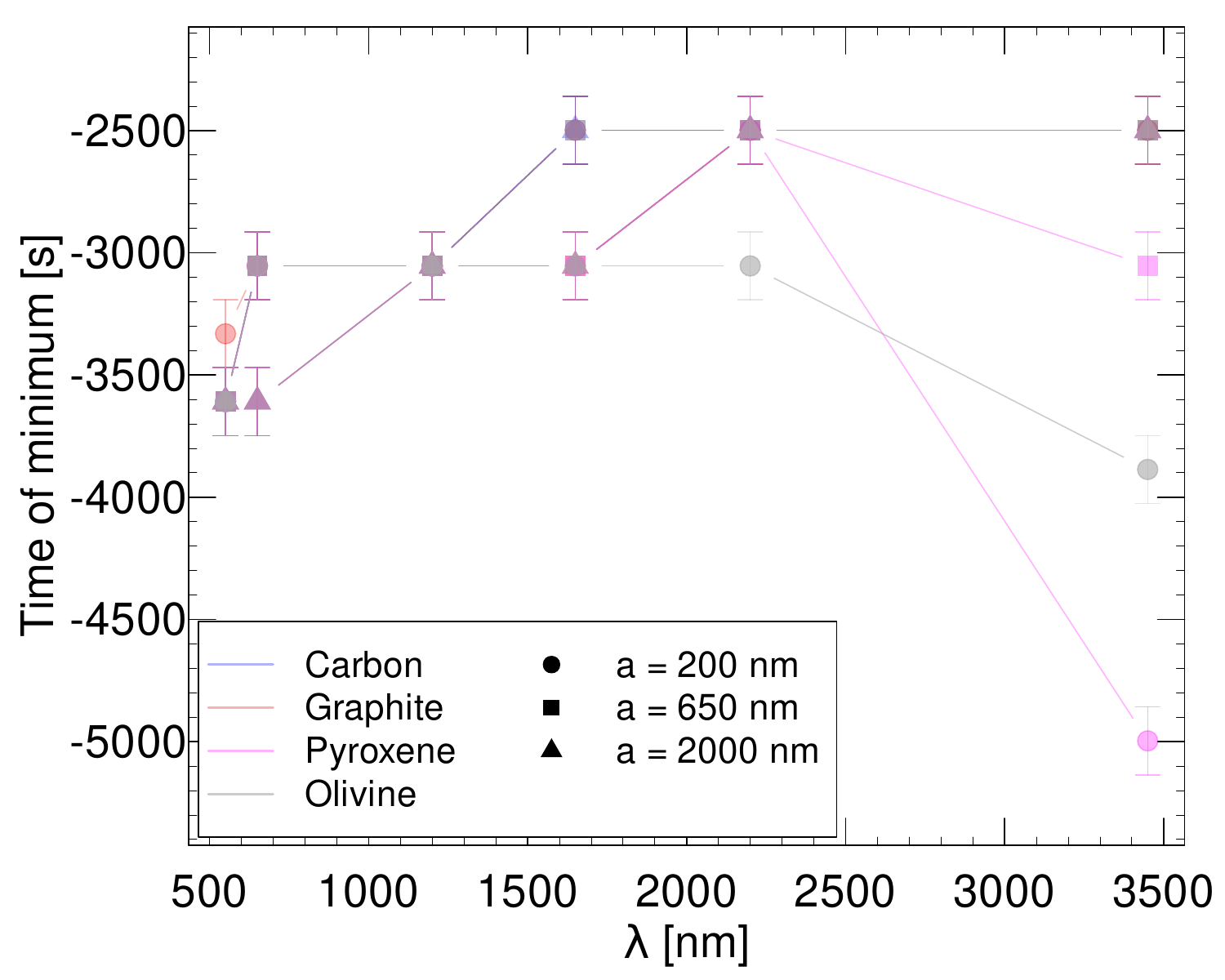}
    \includegraphics[width = .49\textwidth]{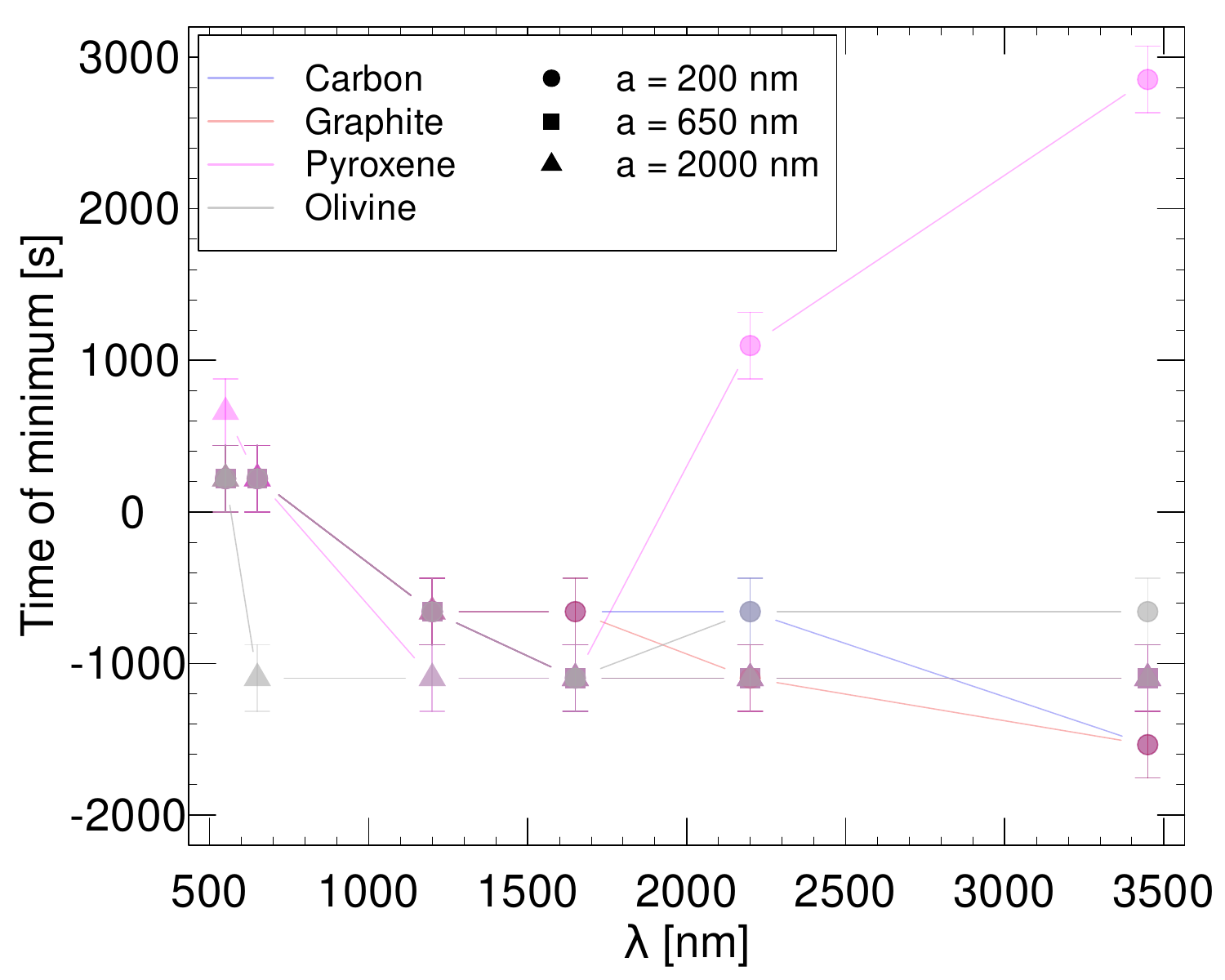}
    \caption{Extracted times of minima in the case of the narrow-tailed comet (Fig. \ref{fig:distributions}, left) for the Sun-like star (left) and the A star (right). Straight lines are shown for visualization purposes only.}
    \label{fig:73p_ttv}
\end{figure}

\begin{figure}
    \centering
    \includegraphics[width = .48\textwidth]{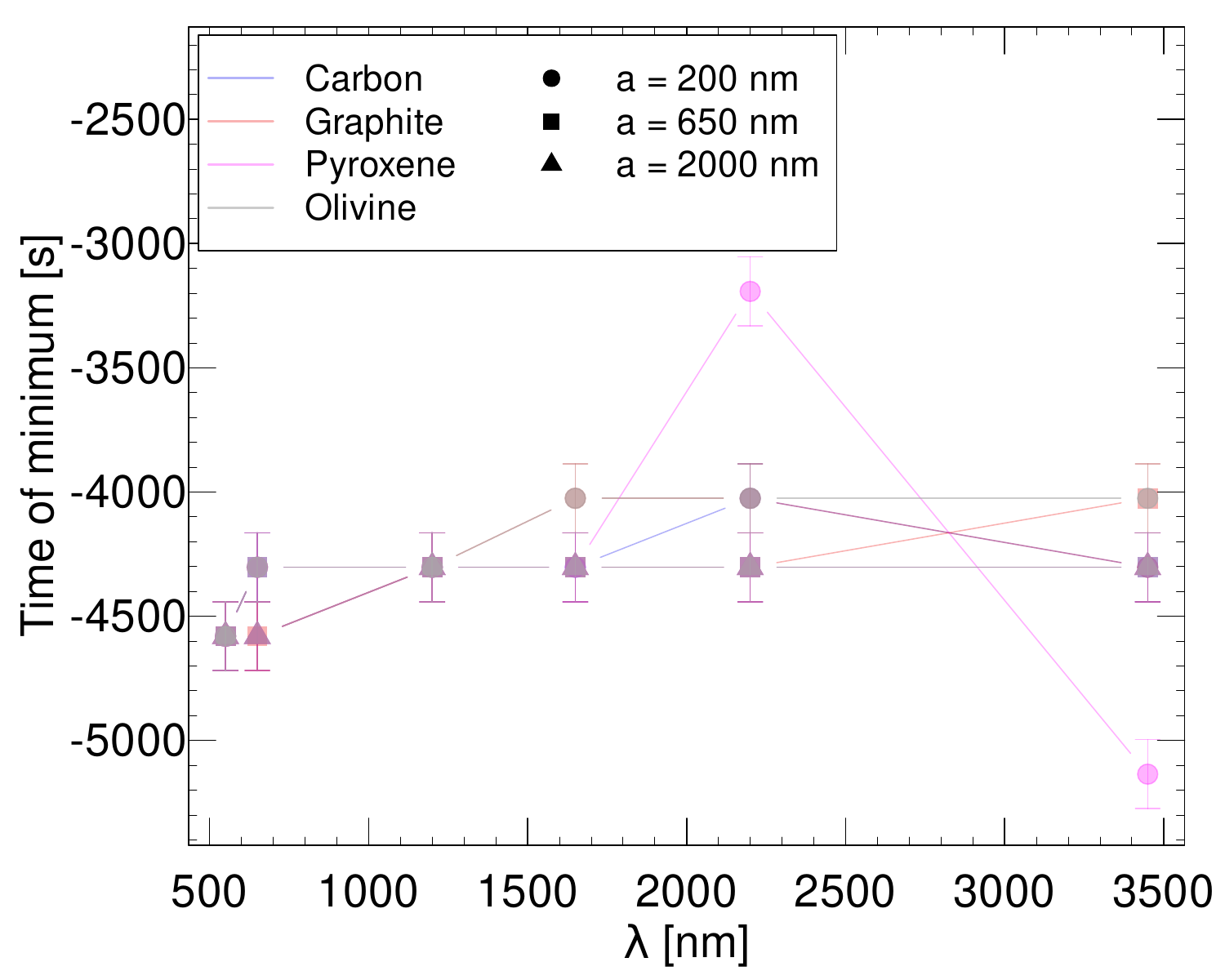}
    \includegraphics[width = .48\textwidth]{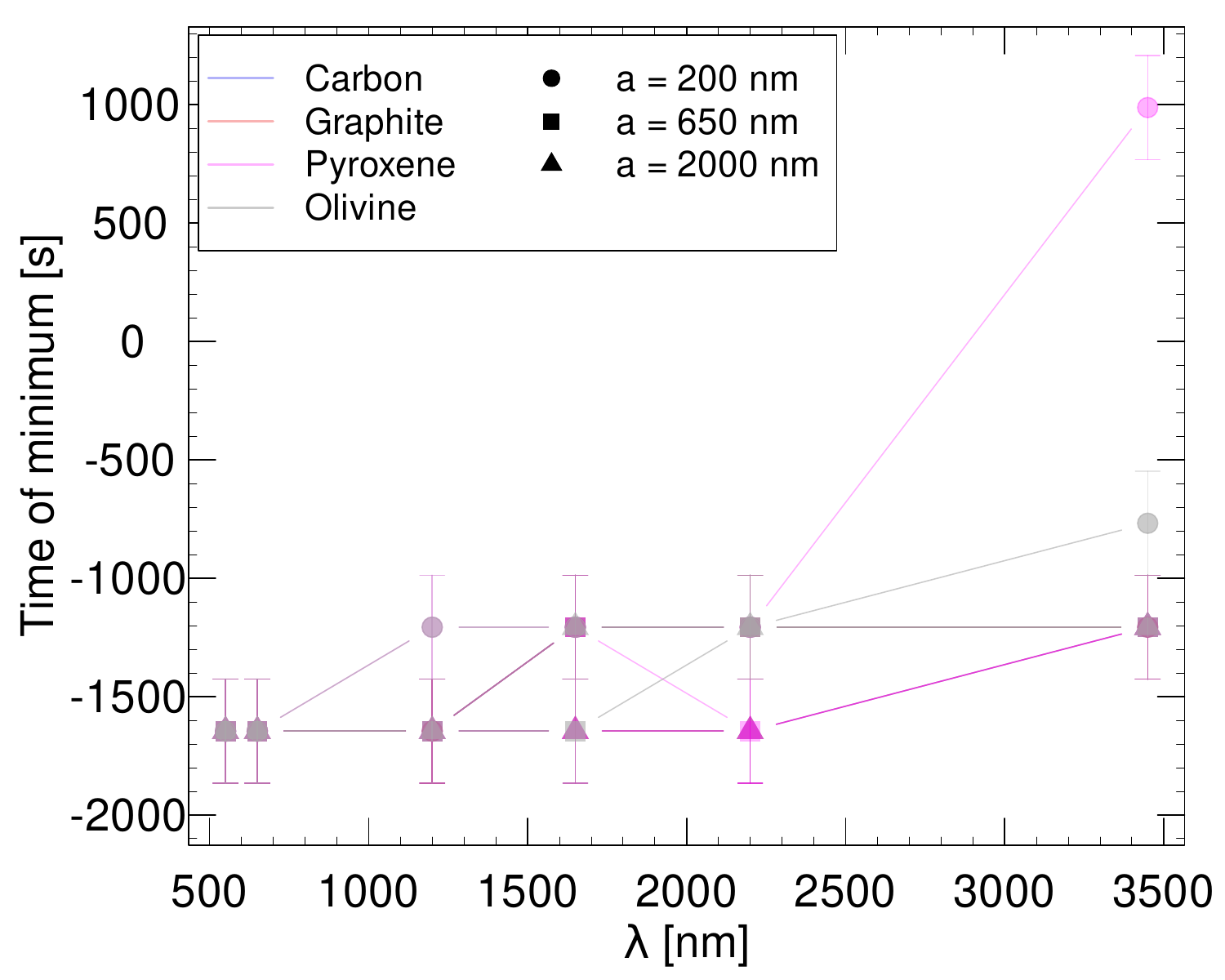}
    \caption{Extracted times of minima in the case of the wide-tailed comet (Fig. \ref{fig:distributions}, middle) for the Sun-like star (left) and the A star (right). Straight lines are shown  for visualization purposes only.}
    \label{fig:29p_ttv}
\end{figure}

We also observe a change in the transit minima of the asymmetric dust distributions (Figs. \ref{fig:73p_v1}, \ref{fig:73p_astar}, \ref{fig:29p_v1}, and \ref{fig:29p_astar}) at different wavelengths. These are analogous to the transit timing variations (TTVs) of exoplanets \citep[e.g.][]{2005MNRAS.359..567A}. By extracting the times of minima (Figs. \ref{fig:73p_ttv} and \ref{fig:29p_ttv}), we show that there may be significant offset at the different wavelengths. We note that $t = 0$ in our simulations corresponds to the collinearity of the observer, the center of dust distribution, and the center of the stellar disk. Due to the asymmetric dust distribution, the minimum of the light curve does not necessarily occur at $t = 0$. This phenomenon occurs because the limb-darkened stellar disk is convolved by the non-spherical comet tails of Figs. \ref{fig:distributions}. We draw caution to the K-- and L--band minima of the pyroxene-comets with $a = 200$\,nm grain sizes, as these show extremely shallow transits, where the determination of the timings may be misleading, although these show the most significant discrepancy with the timing data at other wavelengths. We also note that the transit timing data (Figs. \ref{fig:73p_ttv} and \ref{fig:29p_ttv}) are estimated on the simulation grid (thus, only specific values can be recovered). The uncertainties are likewise estimated from the sampling on the grid.

\subsection{Application to further surveys}

The idea we present here is the multiband observation of nonperiodic photometric transients in stars with FEBs and/or dippers, and the evaluation of the observations in the light of AD ratios before, after and during the transients. There will likely be no unique solution for the multiband signals because the dust scattering scenario is complex and it has degenerations between the dust properties, the size distribution, and the total mass of dust in the line of sight. However, changing colors can still be observed and in favorable cases, the shift of the characteristic dust size can be proven. Here we review a few surveys where the multiband observations or multiband follow-ups promise a good possibility for an exocomet program.

\paragraph{TESS}
Most of the known exocomet transits to date have been discovered by \texttt{TESS} \citep{2015JATIS...1a4003R} in the $\beta$ Pic system \citep{2019A&A...625L..13Z, 2022NatSR..12.5855L}. Given the apparent aperiodic or quasiperiodic nature of the transits themselves, continuous, survey-type photometry over long baselines appears to be the favored strategy. \texttt{TESS} has the advantage of long time series with almost continuous coverage. This is almost a compulsory requirement in ``exocomet hunting'', because the transients occur unpredictably and have a duration in the order of 0.5-1 day -- making them very hard to detect from the ground, most promisingly with an observatory network. \texttt{TESS} represents the largest space photometry factory today, which makes it a prime instrument in exocomet search as well. However, TESS is a single-band photometer and, therefore, a targeted follow-up observation is required for multiband photometry purposes. These follow-up observations may have a longer cadence than the TESS data: space photometry shows the shape of the transient with good precision, while the follow-up observations have the task of measuring precise colors and confirm the presence of possible color variations, where the cadence can be longer than for the ``transit shape'' photometry.

\paragraph{CHEOPS}
Well-tailored observations \citep[with e.g. \texttt{CHEOPS};][]{2021ExA....51..109B} of individual stellar systems can also yield interesting results. A transient feature, related to the debris disk surrounding DE Bootis, has been identified in its \texttt{CHEOPS} light curve by \cite{2023A&A...671A.127B}. A possible exocometary transit was identified in the HD 172555 system using its available \texttt{ CHEOPS} observations \citep{2023A&A...671A..25K}.
Here, synchronous follow-up observations with multiband instruments are also required to map the colors during the transients.

\paragraph{PLATO}
With the 2026 launch of \texttt{PLATO} \citep{2014ExA....38..249R}, offering up to more than a year of uninterrupted ultra-high precision multicolor \citep{2020ExA....50....1G} light curves in the Long-duration Observation Phases \citep{2022A&A...658A..31N}, further detection of these transits is expected.

\paragraph{ARIEL}
The upcoming \texttt{Ariel} mission \citep{2021arXiv210404824T} will be observing individual exoplanet systems from a target list. Ariel is planned to have the ability of synchronous visual + infrared photometry \citep[VISPhot, FGS1 and FGS2;][]{2022ExA....53..607S} together with NIR spectra taken by two low-resolution spectrographs \citep[NIRSpec and AIRS][]{2021arXiv210404824T} with coverage up to $7.8$~$\mu$m, which is also capable of synthetic spectrophotometry. \texttt{Ariel} will be the first dedicated exoplanet space observatory with multiband abilities, and will be very much suited for a confirmation of the comet nature of dusty transients. Moreover, the low-resolution spectra from Ariel can indeed reveal the chemical constituents of the dusty transients, providing an important constraint to the deeper modeling and confirming the origin of them. 

\paragraph{Rubin LSST}
\texttt{LSST} will be a large wide-field ground-based survey designed to obtain repeated images that cover the entire southern celestial hemisphere and the Ecliptic. The 8.4 m telescope of the Rubin Vera Observatory will be equipped with a 3.2-gigapixel camera, and six filters (ugrizy) covering the wavelength range 320-1050 nm \citep{2019ApJ...873..111I}. \texttt{LSST} is not designed to collect photometric data series and, therefore, is not considered as an instrument to discover dippers or proven exocomet transits. Rather, it will be able to follow-up up the already known dippers of the southern sky, providing well calibrated 6 band photometry for 10 years of planned operation, with a pace of $\sim 4$ days. This can be combined with the uninterrupted space photometry of the dippers, revealing the transients with very good time resolution. The combined dataset will be very valuable to study practically any dipper in the southern sky together with the subtle changes of dust color over a very long time basis. This will be an unprecedented opportunity to point out the changes of dust properties during the transient events which are compatible with exocomets, providing very strong evidence to this emerging scientific field.

\section{Summary and concluding remarks} \label{sec:disc}

The conclusions of this article can be summarized in the following points.

\begin{itemize}
    \item 
We have introduced an application based on RADMC-3D that allows simulating the transits of 3D, $256^3$ pixelated comet models in front of stars at any photometric wavelength, assuming an arbitrary distance from the star, an arbitrary dust composition, and wavelength of observation.

    \item 
Inspired by the comets in the Solar System, we calculated transit models assuming compositions of carbonaceous (carbon and graphite) and silicate (pyroxene and olivine) dust, at the central wavelength of the VRJHKL band for three comet models with a thin tail, a fan-like tail, and no tail.

    \item 
The behavior of silicate and carbonaceous comet dust proved to be distinguishable in simulated VRJKHL transit observations. We also successfully simulated the dependence of wavelength-scattering cross-section curves on particle sizes.

    \item 
The origin of once-frozen comet dust in the dusty environment of extrasolar systems can be discerned from the hot/warm circumstellar dust mostly due to the differing size distribution, and the cometary nature can be confirmed through the different particle distributions and chemical compositions. Synchronuous optical+NIR observations are well suited for this task.

    \item 
We presented some possible applications in planned space telescope missions in the present and near future.
\end{itemize}

Observational astrophysics of dippers and systems showing the FEB phenomenon and investigating possible exocomets in these systems are an emerging area in extrasolar astronomy. Our results demonstrated the potential of multiband observations in this field and point toward promising opportunities.

\begin{acknowledgments}
The project has been implemented with the support provided by the Ministry of Culture and Innovation of Hungary
from the National Research, Development and Innovation Fund, financed under the SNN-147362 funding scheme.
This work was partly supported by the K-138962 grant and the TKP2021-NKTA-64 excellence grant of the National Research, Development and Innovation Office (NKFIH, Hungary). SzK and GyMSz thank the support of the PRODEX Experiment Agreement No. 4000137122 between the ELTE E\"otv\"os Lor\'and University and the European Space Agency (ESA-D/SCI-LE-2021-0025).This research has made use of the NASA Exoplanet Archive, which is operated by the California Institute of Technology, under contract with the National AeronAUtics and Space Administration under the Exoplanet Exploration Program. Project no. C1746651 has been implemented with the support provided by the Ministry of Culture and Innovation of Hungary from the National Research, Development and Innovation Fund, financed under the NVKDP-2021 funding scheme. The generated light curves will be made publicly available upon acceptance of the paper at \url{https://osf.io/ndjw6/?view_only=d9f0e84412a14e169e3de044628be6ab}.

\end{acknowledgments}

%

\vspace{5mm}
\facilities{Exoplanet Archive}


\software{RADMC-3D \citep{radmc}, optool \citep{optool}}




\bibliography{sample631}{}
\bibliographystyle{aasjournal}



\appendix
\section{Light curves for a Sun-like star}\label{sect:app_lc}

\begin{figure}[!h]
    \centering
    \includegraphics[width = \textwidth]{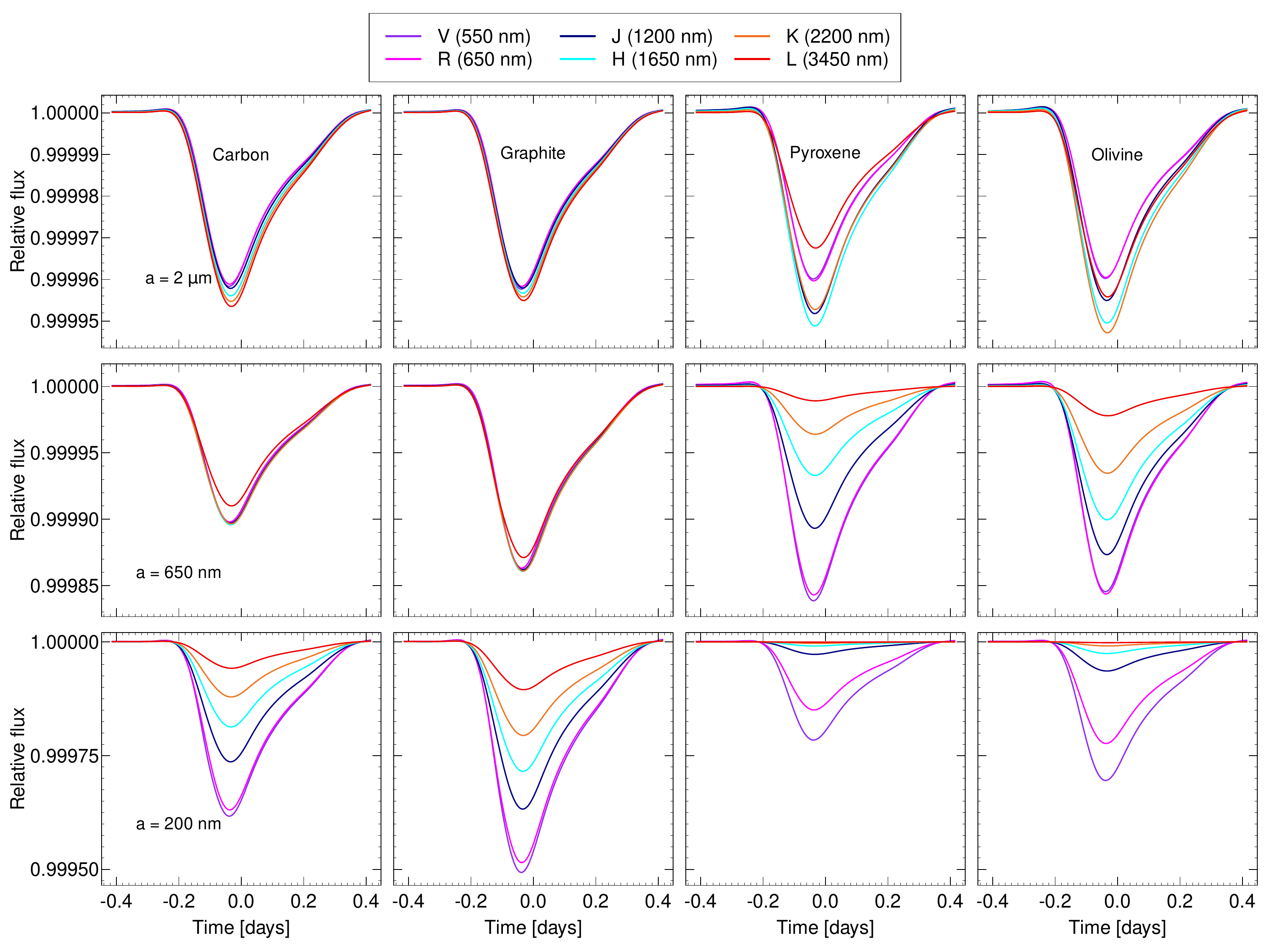}
   \caption{Same as Fig. \ref{fig:73p_astar} but the transits are calculated for solar-like star.}
    \label{fig:73p_v1}
\end{figure}

\begin{figure}
    \centering
    \includegraphics[width = \textwidth]{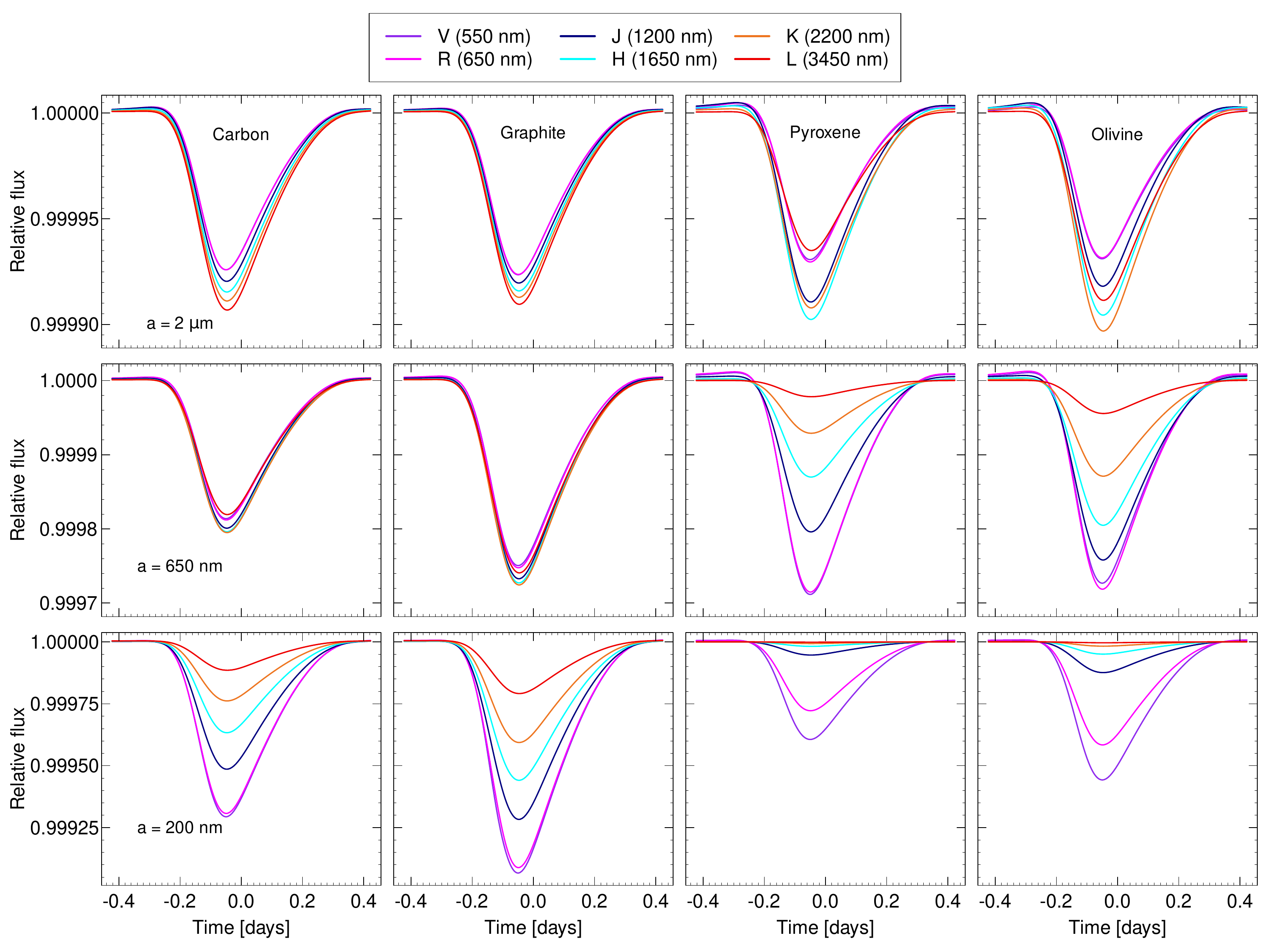}
 \caption{Same as Fig. \ref{fig:29p_astar} but the transits are calculated for Solar-like star.}    
 \label{fig:29p_v1}
\end{figure}

\begin{figure}
    \centering
    \includegraphics[width = \textwidth]{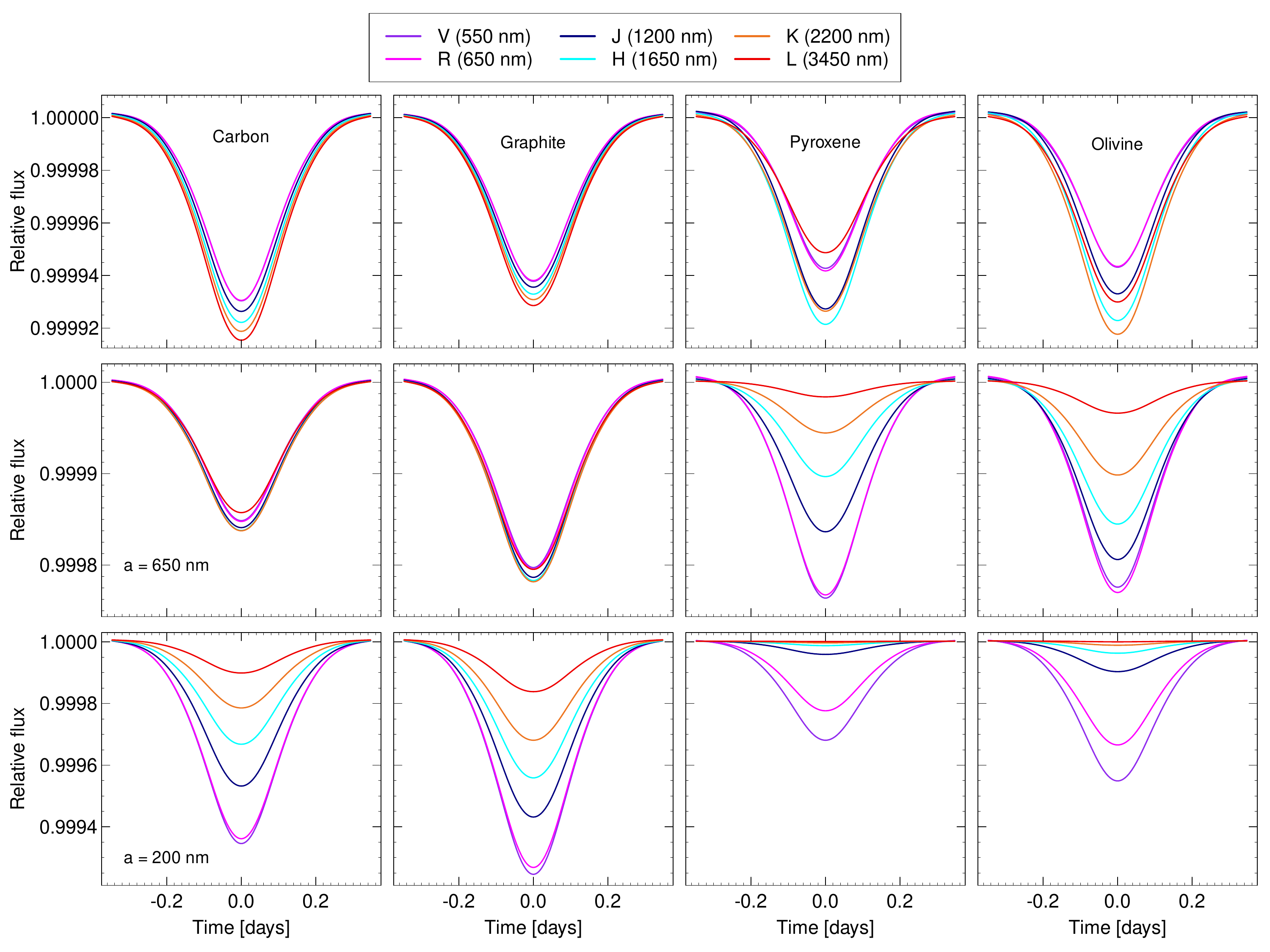}
\caption{Same as Fig. \ref{fig:17p_astar} but the transits are calculated for a Solar-like star.}    \label{fig:17p_v1}
\end{figure}

In this section, we show the light curves calculated in Sections \ref{sec:res} (Figs. \ref{fig:73p_astar}
--\ref{fig:17p_astar}), but for a Solar-like host star (Figs. \ref{fig:73p_v1}
--\ref{fig:17p_v1}). The light curves are morphologically similar, whereas quantitative deviations occur due to the differing stellar radius and the differing limb darkening (see its discussion in \ref{sect:lc_calc}).

\clearpage
\section{Color variations}\label{sect:app_colvar}

In this section, we plot the differential light curves calculated in different simulation bandpasses. These figures show the color of the total stellar light after having traveled through the comet dust and suffered a spatially dependent transfer, including both outward and inward scattering. The figures are plotted in units of $\rm \mu$mag, and the $V-X$ colors indicate a redder color towards positive values (see its discussion in Sect. \ref{sect:lc_calc}.

\begin{figure}
    \centering
    \includegraphics[width = \textwidth]{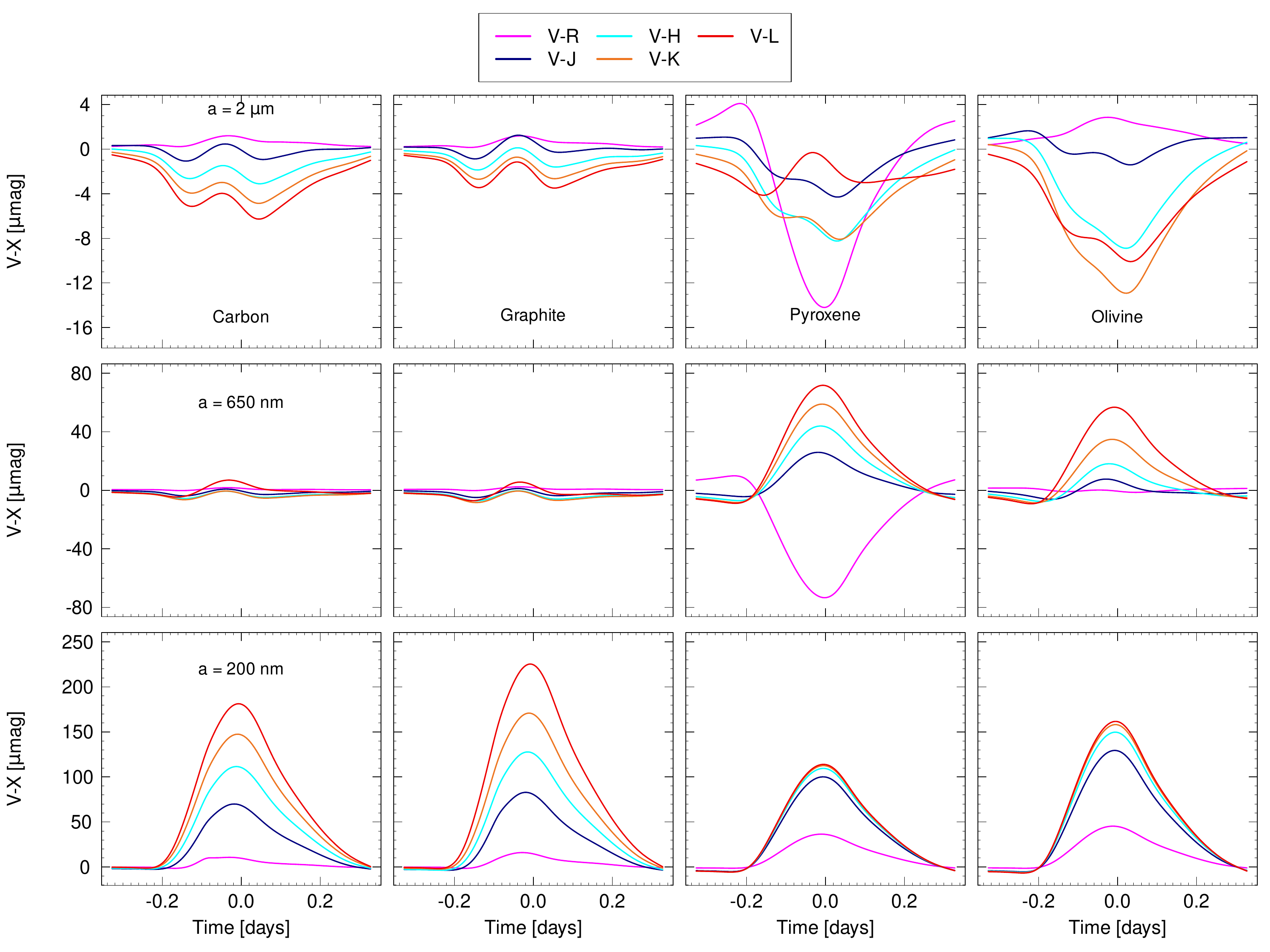}
    \caption{Color variations for the narrow-tailed comet (Fig. \ref{fig:distributions}) orbiting an A star.}
    \label{fig:73p_astar_colors}
\end{figure}

\begin{figure}
    \centering
    \includegraphics[width = \textwidth]{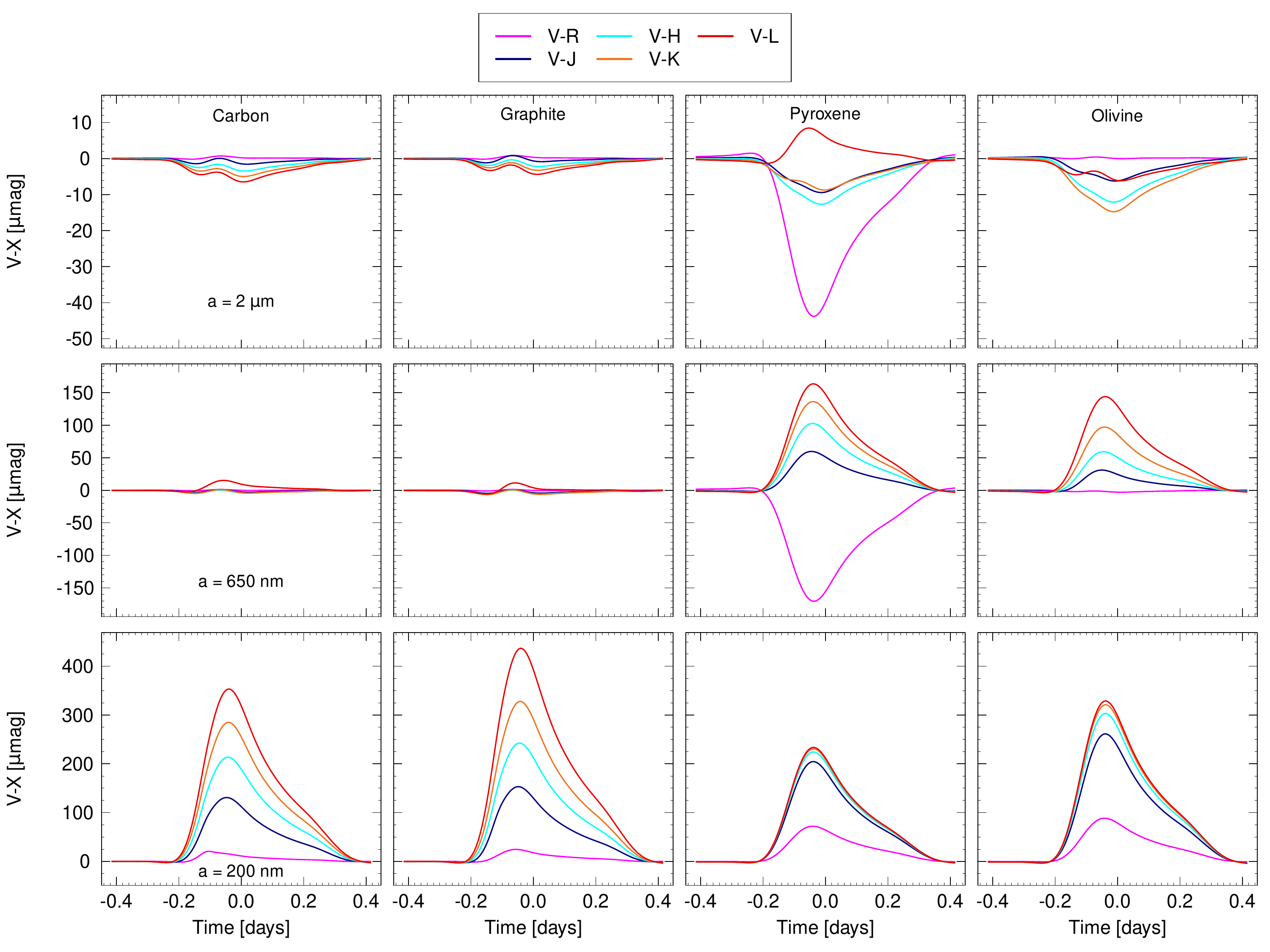}
    \caption{Color variations for the narrow-tailed comet (Fig. \ref{fig:distributions}) orbiting a solar-like star.}
    \label{fig:73p_solar_colors}
\end{figure}

\begin{figure}
    \centering
    \includegraphics[width = \textwidth]{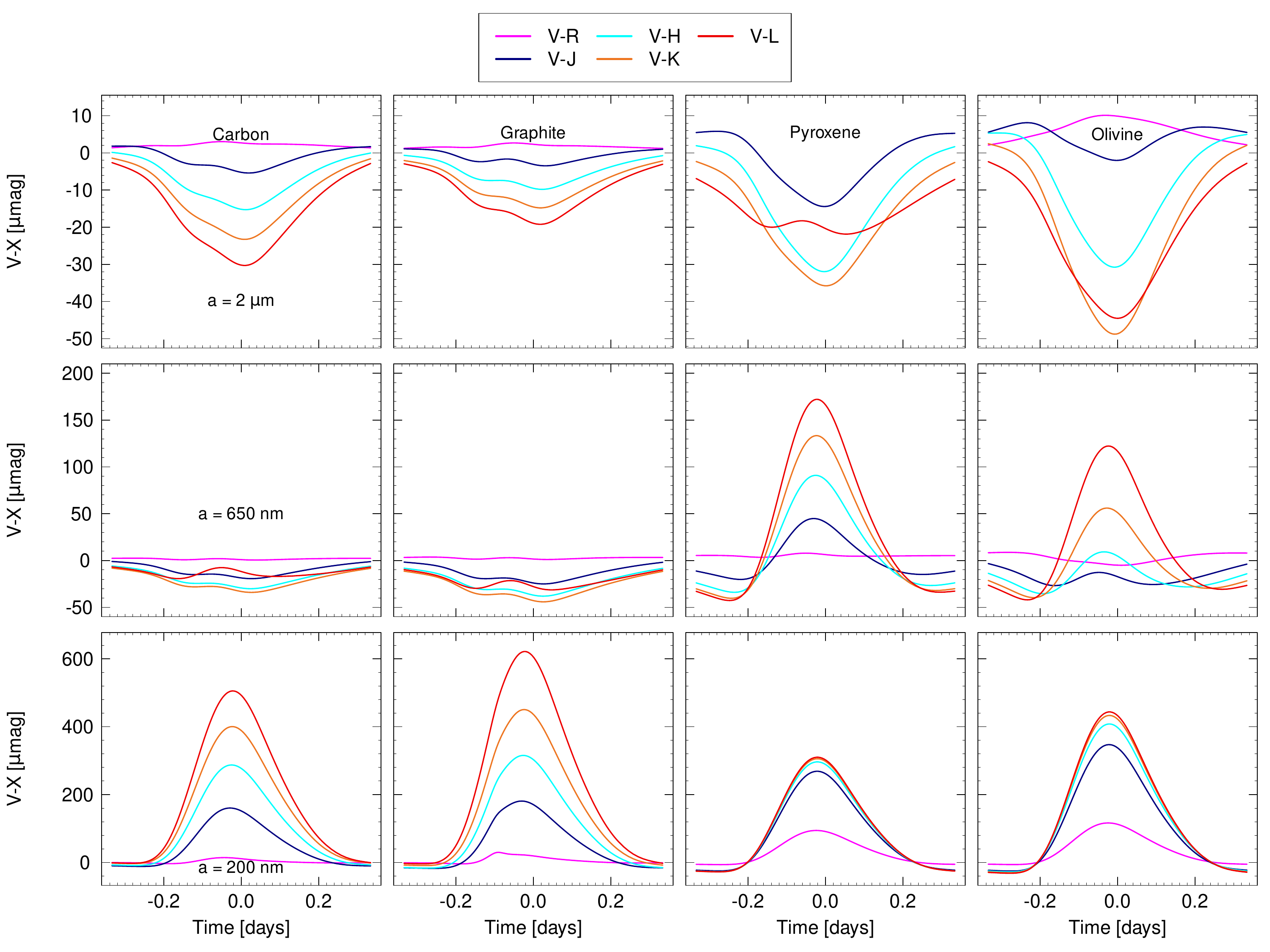}
    \caption{Color variations for the wide-tailed comet (Fig. \ref{fig:distributions}) orbiting an A star.}
    \label{fig:29p_astar_colors}
\end{figure} 

\begin{figure}
    \centering
    \includegraphics[width = \textwidth]{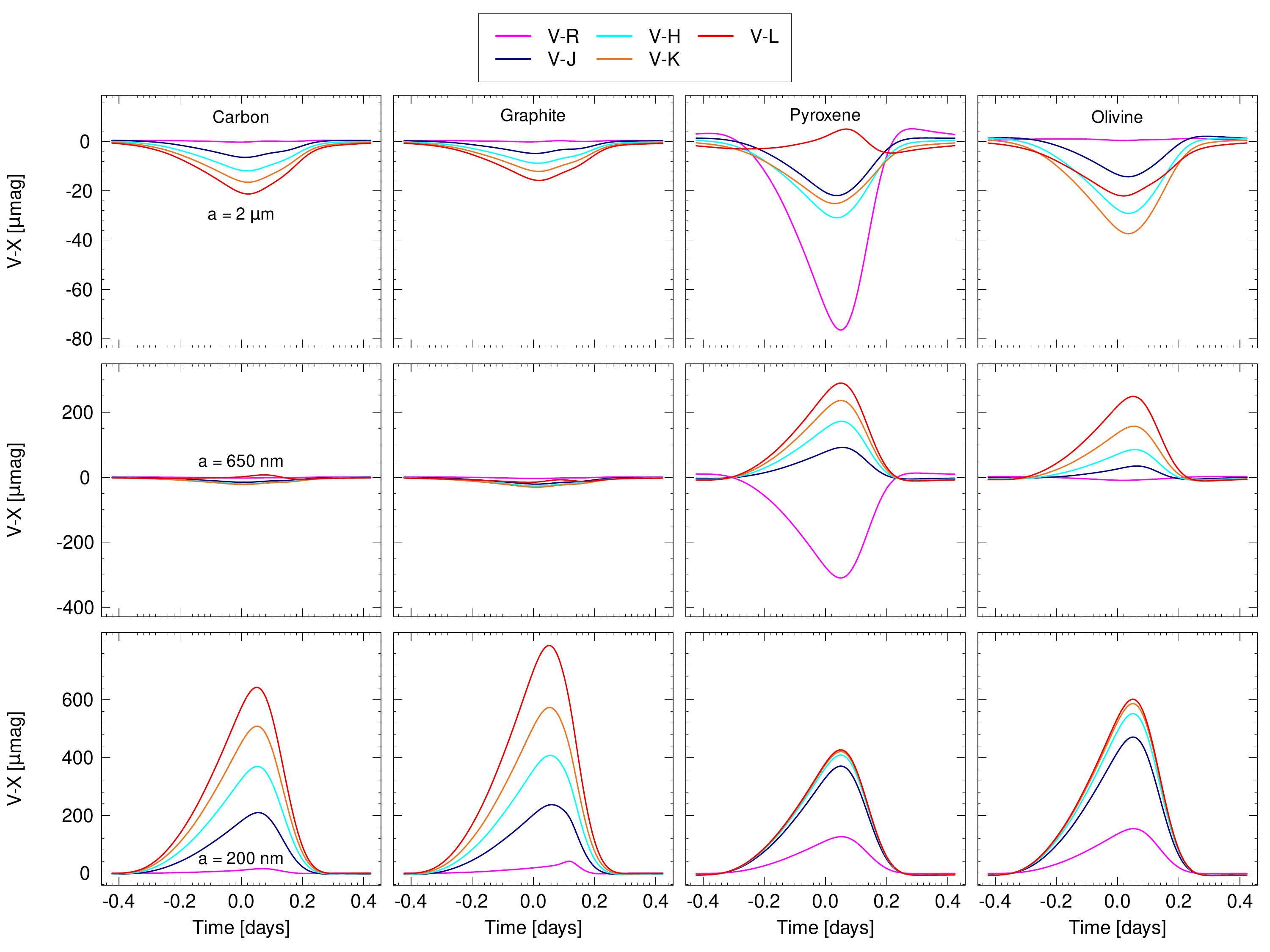}
    \caption{Color variations for the wide-tailed comet (Fig. \ref{fig:distributions}) orbiting a solar-like star.}
    \label{fig:29p_solar_colors}
\end{figure}

\begin{figure}
    \centering
    \includegraphics[width = \textwidth]{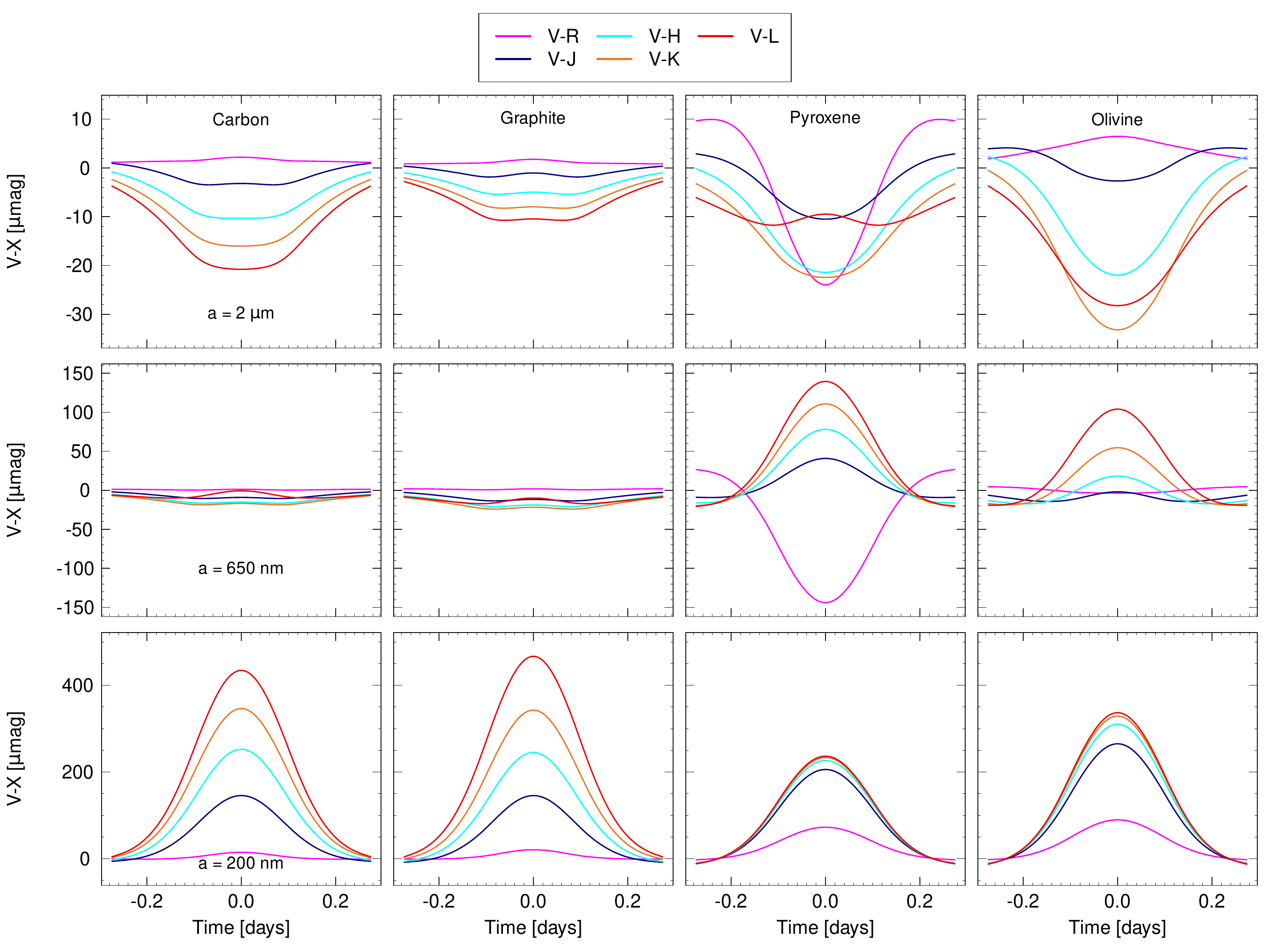}
    \caption{Color variations for spherically symmetric comet (\ref{fig:distributions}) orbiting an A star.}
    \label{fig:h2_astar_colors}
\end{figure}
\begin{figure}
    \centering
    \includegraphics[width = \textwidth]{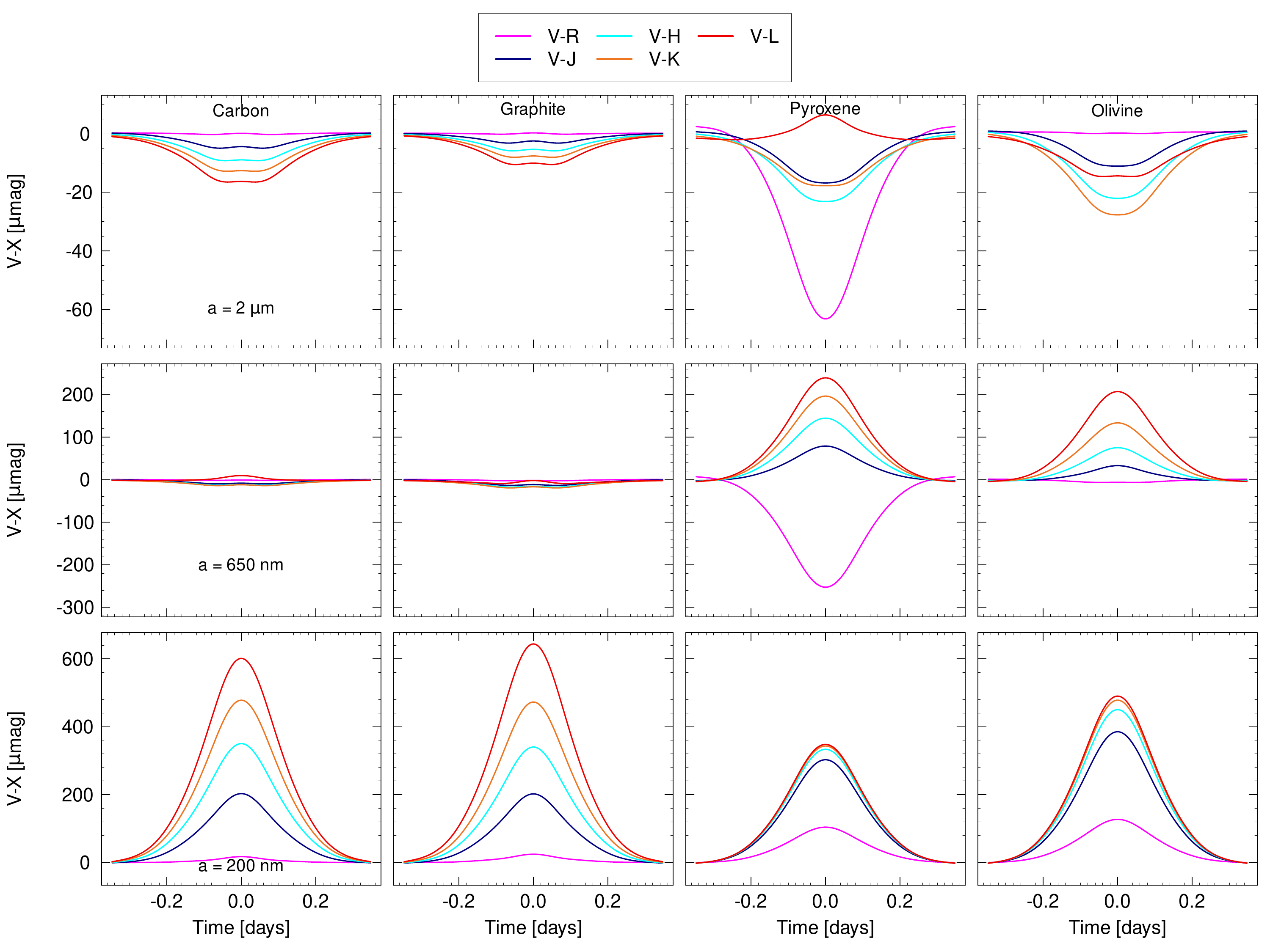}
    \caption{Color variations for spherically symmetric comet (Fig. \ref{fig:distributions}) orbiting a solar-like star.}
    \label{fig:h2_solar_colors}
\end{figure}

\end{document}